\documentclass[preprint,nofootinbib]{revtex4}%
\usepackage{amssymb}
\usepackage{amsfonts}
\usepackage{amsmath}
\usepackage{amsmath}
\usepackage{graphicx}
\usepackage[usenames]{color}%
\usepackage{hyperref}
\providecommand{\U}[1]{\protect\rule{.1in}{.1in}}
\providecommand{\U}[1]{\protect\rule{.1in}{.1in}}
\definecolor{blue}{rgb}{0,0,1}

\definecolor{red}{rgb}{1,0,0}

\begin{document}
\title{New BPS solitons in $\mathcal{N}=4$ gauged supergravity and black holes in Einstein-Yang-Mills-dilaton theory}

\author{Fabrizio Canfora}
\affiliation{Centro de Estudios Científicos (CECs), Casilla 1469, Valdivia, Chile}

\author{Julio Oliva and Marcelo Oyarzo}
\affiliation{Departamento de Física, Universidad de Concepción, Casilla 160-C, Concepción, Chile}

\begin{abstract}
We start by revisiting the problem of finding BPS solutions in
$\mathcal{N}=4$ $SU\left(  2\right)  \times SU\left(  2\right)  $ gauged
supergravity. We report on a new supersymmetric solution in the Abelian sector of the theory, which describes a soliton that is regular everywhere. The solution is 1/4 BPS and can be obtained from a double analytic continuation of a planar solution found by Klemm in hep-th/9810090. Also in the Abelian sector, but now for a spherically symmetric ansatz we find a new solution whose supersymmetric nature was overlooked in the previous literature.
Then, we move to the non-Abelian sector of the theory by considering the meron
ansatz for $SU\left(  2\right)  $. We construct electric-meronic and
double-meron solutions and show that the latter also leads to 1/4 BPS
configurations that are singular and acquire an extra conformal Killing vector. We then move beyond the supergravity embedding of this theory by
modifying the self-interaction of the scalar, but still within the same meron
ansatz for a single gauge field, which is dilatonically coupled with the
scalar. We construct exact black holes for two families of self-interactions
that admit topologically Lifshitz black holes, as well as other black holes
with interesting causal structures and asymptotic behavior. We analyze some
thermal properties of these spacetimes.

\end{abstract}
\maketitle

\section{Introduction}

The Einstein-Maxwell-dilaton theory is a simple setup that allows to avoid
black hole no-hair results \cite{Gibbons:1987ps}. Besides of being an
interesting model in its own right to explore black hole physics, this theory
is relevant since it naturally emerges from the low energy limit of String
Theory, as well as from the dimensional reduction of General Relativity from
five to four dimensions. General Relativity with scalar fields non-minimally
coupled to non-Abelian gauge fields also naturally emerge in gauged
supergravity theories, in which the dynamics of the scalars is supplemented by
a self-interaction. It is well-known that gauged supergravities may possess AdS
vacua located at the extrema of the potential for the scalars. Nevertheless,
there are cases in which the potential does not have an extremum to support a
maximally symmetric vacuum. One of such cases is the Freedman and Schwarz
theory, that corresponds to the four-dimensional, $\mathcal{N}=4$ $SU\left(
2\right)  \times SU\left(  2\right)  $ gauged supergravity \cite{FS}. Even though no
maximally symmetric vacuum exists in this theory, there are electrically and
magnetically charged solutions in which the spacetime is still highly
symmetric and given by the products $AdS_{2}\times S^{2}$ or $AdS_{2}\times
\mathbb{R}^{2}$, where the latter might have 1/4 or 1/2 unbroken supersymmetries \cite{GF}. The Freedman-Schwarz model was later recognized as
a dimensional reduction of $\mathcal{N}=1$, 10-dimensional SUGRA, when the
compactification is carried out on $S^{3}\times S^{3}$ \cite{CVcorto,CVlargo}, and therefore, the model can also be embedded in 11-dimensional supergravity \cite{onceadiez}.

The field content of $\mathcal{N}=4$ $SU\left(  2\right)  \times SU\left(
2\right)  $ gauged supergravity is given by a vielbein $e_{\mu}^{\ a}$, four
Majorana spin-3/2 fields $\psi_{\mu}\equiv\psi_{\mu}^{I}$, as well as four
Majorana spin-1/2 fields $\chi\equiv\chi^{I}$, where the index $I=1,2,3,4,$
runs in the fundamental of $SU\left(  2\right)  \times SU\left(  2\right)  $.
The theory also contains a pseudoscalar axion field $\mathbf{a}(x)$, as well as a
real scalar $\phi(x)$, namely the dilaton. The Yang-Mills sector consists of a
vector and a pseudovector, non-abelian gauge field $A_{\ \mu}^{i}$ and
$B_{\ \mu}^{i}$, respectively, with independent gauge couplings $e_{A}$ and
$e_{B}$, where the index $i=1,2,3$ transforms in the adjoint of each corresponding $SU\left(
2\right)$ copy. Following the conventions of \cite{Klemm}, the
action reads%
\begin{align}
\frac{\mathcal{L}}{\sqrt{-g}}  &  =-\frac{R}{4}+\frac{1}{2}\left[  \left(
\partial\phi\right)  ^{2}+e^{4\phi}\left(  \partial\mathbf{a}\right)
^{2}\right]  -V\left(  \phi\right)  -\frac{e^{-2\phi}}{4}\left(  A^{i\mu\nu
}A_{i\mu\nu}+B^{i\mu\nu}B_{i\mu\nu}\right) \nonumber\\
&  -\frac{\mathbf{a}}{2}\left(  \tilde{A}_{i\mu\nu}A^{i\mu\nu}+\tilde{B}%
_{i\mu\nu}B^{i\mu\nu}\right)  \ , \label{Lsugrafull}%
\end{align}
where $\sqrt{-g}=e=\det\left(  e_{\mu}^{\ a}\right)  $ and the
self-interaction of the dilaton is unbounded from below and it is given by%
\begin{equation}
V\left(\phi\right)=-\frac{\left(  e_{A}^{2}+e_{B}^{2}\right)  }{8}e^{2\phi}\ , \label{Vsugra}%
\end{equation}
As usual%
\begin{align}
A_{i\mu\nu}  &  =\partial_{\mu}A_{\ \nu}^{i}-\partial_{\nu}A_{\ \mu}^{i}%
+e_{A}\epsilon_{ijk}A_{\ \mu}^{j}A_{\ \nu}^{k}\ ,\\
B_{i\mu\nu}  &  =\partial_{\mu}B_{\ \nu}^{i}-\partial_{\nu}B_{\ \mu}^{i}%
+e_{B}\epsilon_{ijk}B_{\ \mu}^{j}B_{\ \nu}^{k}\ ,\\
\tilde{A}_{\ \mu\nu}^{i}  &  =\frac{1}{2\sqrt{-g}}\epsilon_{\mu\nu\rho\sigma
}A^{i\rho\sigma}\ .
\end{align}
The supersymmetry transformations are generated by the local spinorial parameter $\epsilon\left(
x\right)  \equiv\epsilon^{I}\left(  x\right)  $, which on a purely bosonic
configuration, when acting on the fermionic fields of the theory reduce to%
\begin{align}
\delta\bar{\chi}  &  =\frac{i}{\sqrt{2}}\bar{\epsilon}\left(
\partial_{\mu}\phi+i\gamma_{5}e^{2\phi}\partial_{\mu}\mathbf{a}\right)
\gamma^{\mu}-\frac{1}{2}e^{-\phi}\bar{\epsilon}C_{\mu\nu}\sigma^{\mu\nu}%
+\frac{1}{4}e^{\phi}\bar{\epsilon}\left(  e_{A}+i\gamma_{5}e_{B}\right)
\ ,\label{unmedio-transf}\\
\delta\bar{\psi}_{\rho}  &  =\bar{\epsilon}\left(  \overleftarrow
{D}_{\rho}-\frac{i}{2}e^{2\phi}\gamma_{5}\partial_{\rho}\mathbf{a}\right)
-\frac{i}{2\sqrt{2}}e^{-\phi}\bar{\epsilon}C_{\mu\nu}\gamma_{\rho}\sigma
^{\mu\nu}+\frac{i}{4\sqrt{2}}e^{\phi}\bar{\epsilon}\left(  e_{A}+i\gamma
_{5}e_{B}\right)  \gamma_{\rho}\ . \label{RS-transf}%
\end{align}
The Lorentz and gauge covariant derivative is given by%
\begin{equation}
\overleftarrow{D}_{\rho}=\overleftarrow{\partial}_{\rho}-\frac{1}{2}%
\omega_{\rho}^{\ ab}\sigma_{ab}+\frac{1}{2}e_{A}\alpha^{i}A_{\ \rho}^{i}%
+\frac{1}{2}e_{B}\beta^{i}B_{\rho}^{i}\ ,
\end{equation}
and the generators we will use are given by the following $4\times4$ matrices
(see \cite{FS})%
\begin{align}
\alpha^{1}  &  =\left(
\begin{array}
[c]{cc}%
0 & \sigma_{1}\\
-\sigma_{1} & 0
\end{array}
\right)  \ ,\qquad\quad\alpha^{2}=\left(
\begin{array}
[c]{cc}%
0 & -\sigma_{3}\\
\sigma_{3} & 0
\end{array}
\right)  \ ,\qquad\quad\alpha^{3}=\left(
\begin{array}
[c]{cc}%
i\sigma_{2} & 0\\
0 & i\sigma_{2}%
\end{array}
\right)  \ ,\label{losalphas}\\
\beta^{1}  &  =\left(
\begin{array}
[c]{cc}%
0 & -i\sigma_{2}\\
-i\sigma_{2} & 0
\end{array}
\right)  \ ,\qquad \beta^{2}=\left(
\begin{array}
[c]{cc}%
0 & -1\\
1 & 0
\end{array}
\right)  \ ,\qquad\quad\,\beta^{3}=\left(
\begin{array}
[c]{cc}%
i\sigma_{2} & 0\\
0 & -i\sigma_{2}%
\end{array}
\right)  \ .\label{losbetas}\\
\sigma_{1}  &  =\left(
\begin{array}
[c]{cc}%
0 & 1\\
1 & 0
\end{array}
\right)  ,\qquad\sigma_{2}=\left(
\begin{array}
[c]{cc}%
0 & -i\\
i & 0
\end{array}
\right)  ,\qquad\sigma_{3}=\left(
\begin{array}
[c]{cc}%
1 & 0\\
0 & -1
\end{array}
\right) \nonumber
\end{align}
which generate the algebra $su\left(  2\right)  \times su\left(  2\right)  $,
namely,%
\begin{align}
\alpha^{i}\alpha^{j}  &  =-\delta^{ij}-\epsilon^{ijk}\alpha^{k}\ ,\qquad
\quad\beta^{i}\beta^{j}=-\delta^{ij}-\epsilon^{ijk}\beta^{k}\ ,\qquad
\quad\left[  \alpha^{i},\beta^{j}\right]  =0\ .\\
\left[  \alpha^{i},\alpha^{j}\right]   &  =-2\epsilon^{ijk}\alpha^{k}%
\ ,\qquad\left[  \beta^{i},\beta^{j}\right]  =-2\epsilon^{ijk}\beta
^{k}\ .\qquad
\end{align}
Furthermore, following \cite{FS} we have defined%
\begin{equation}
C_{\mu\nu}\equiv\alpha^{i}A_{\ \mu\nu}^{i}+i\gamma_{5}\beta^{i}B_{\ \mu\nu
}^{i}\ .
\end{equation}
As it is well known, in the absence of torsion the spin connection is given by%
\begin{equation}
\omega_{\mu}^{\ ab}=-\frac{1}{2}e_{\mu c}\left(  \Omega^{abc}+\Omega
^{cba}-\Omega^{cab}\right)  \ ,
\end{equation}
where $\Omega$ are the anholonomity coefficients%
\begin{equation}
\Omega_{ab}^{\ \ c}\equiv e_{\ a}^{\mu}e_{\ b}^{\nu}\left(  \partial_{\mu
}e_{\nu}^{\ c}-\partial_{\nu}e_{\mu}^{\ c}\right)  \ .
\end{equation}
and finally the curvature associated with the spin connection is given by%
\begin{equation}
R_{\mu\nu}^{\ \ ab}\left(  \omega\right)  =\partial_{\mu}\omega_{\nu}%
^{\ ab}-\partial_{\nu}\omega_{\mu}^{\ ab}+\omega_{\mu}^{\ ac}\omega_{\nu
c}^{\ \ b}-\omega_{\nu}^{\ ac}\omega_{\mu c}^{\ \ b}\ \ .
\end{equation}

For the fermionic sector we consider the following conventions: $\left\{
\gamma_{\mu},\gamma_{\nu}\right\}  =2g_{\mu\nu}\ $, $\sigma_{\mu\nu}=\frac
{1}{4}\left[  \gamma_{\mu},\gamma_{\nu}\right]  \equiv\frac{1}{2}\gamma
_{\mu\nu}\ $,\ $\gamma_{5}=-i\gamma_{0}\gamma_{1}\gamma_{2}\gamma_{3}$, so
that $\gamma_{5}^{2}=1$ and $\left\{  \gamma_{5},\gamma_{a}\right\}  =0$. When necessary, we will use the following basis%
\begin{equation}
\gamma_{0}=\left(
\begin{array}
[c]{cc}%
-1 & 0\\
0 & 1
\end{array}
\right)  \ ,\ \gamma_{1}=\left(
\begin{array}
[c]{cc}%
0 & \sigma_{1}\\
-\sigma_{1} & 0
\end{array}
\right)  \ ,\ \gamma_{2}=\left(
\begin{array}
[c]{cc}%
0 & \sigma_{2}\\
-\sigma_{2} & 0
\end{array}
\right)  \ ,\ \gamma_{3}=\left(
\begin{array}
[c]{cc}%
0 & \sigma_{3}\\
-\sigma_{3} & 0
\end{array}
\right)  \ .
\end{equation}
The BPS configurations are such that they lead to non-trivial solutions for
$\epsilon$ from the equation $\delta\chi=0$ and $\delta\psi_{\mu}=0$ in
(\ref{unmedio-transf}) and (\ref{RS-transf}), respectively.

In order to analyze the existence of Killing spinors, it is customary to
consider consistency conditions that emerge from the manipulation of equations (\ref{unmedio-transf}) and (\ref{RS-transf}). In particular, plugging the
equation $\delta\chi=0$ from (\ref{unmedio-transf}), in equation (\ref{RS-transf})
implies that BPS solutions must fulfill%
\begin{equation}
\delta\bar{\psi}_{\rho}=\bar{\epsilon}\left(  \overleftarrow{D}_{\rho}%
-\frac{i}{\sqrt{2}}e^{-\phi}C_{\rho\mu}\gamma^{\mu}+\frac{1}{2}\not \partial
\phi\gamma_{\rho}\right)  -\bar{\epsilon}\frac{i}{2}e^{2\phi}\gamma_{5}\left(
\partial_{\rho}\mathbf{a}-\not \partial \mathbf{a}\gamma_{\rho}\right)  =0\ ,
\end{equation}
where $\not \partial\mathbf{a} := \gamma^{\mu}\partial_{\mu}\mathbf{a}$. An integrability conditions for the
Killing spinor comes from imposing $\delta\bar{\psi}_{[\rho}\overleftarrow
{D}_{\sigma]}=0$\ , which after a lengthy but straightforward computation
leads to%
\begin{align}
&  0=\delta\bar{\psi}_{[\rho}\overleftarrow{D}_{\sigma]}\label{consistency}\\
&  =-\frac{1}{8}\bar{\epsilon}R_{\sigma\rho}^{\ \ ab}\gamma_{ab}+\frac{1}%
{4}\bar{\epsilon}g_{A}\alpha^{i}A_{i\sigma\rho}+\frac{1}{4}\bar{\epsilon}%
g_{B}\beta^{i}B_{i\sigma\rho}+\bar{\epsilon}\frac{1}{2}e^{-2\phi}C_{[\sigma
|}^{\ \nu}\gamma_{\nu}C_{|\rho]\mu}\gamma^{\mu}+\bar{\epsilon}\frac{i}%
{\sqrt{2}}e^{-\phi}\partial_{\lbrack\sigma}\phi C_{\rho]\mu}\gamma^{\mu
}\nonumber\\
&  +\bar{\epsilon}\frac{i}{\sqrt{2}}e^{-\phi}\left[  -\alpha^{i}\left(
\nabla_{\lbrack\sigma}A_{\rho]\nu}^{i}+g_{A}\epsilon^{lji}A_{\ \rho\nu}%
^{j}A_{\ \sigma}^{l}\right)  \gamma^{\nu}-i\beta^{i}\gamma_{5}\left(
\nabla_{\lbrack\sigma}B_{\rho]\nu}^{i}+g_{B}\epsilon^{lji}B_{\ \rho\nu}%
^{j}B_{\ \sigma}^{l}\right)  \gamma^{\nu}\right] \nonumber\\
&  +\bar{\epsilon}\frac{1}{2}\nabla_{\lbrack\sigma|}\partial_{\mu}\phi
\gamma^{\mu}\gamma_{|\rho]}+\bar{\epsilon}\frac{1}{4}\partial_{\mu}%
\phi\partial^{\mu}\phi\gamma_{\sigma\rho}-\bar{\epsilon}\frac{1}%
{2}\not \partial \phi\partial_{\lbrack\sigma}\phi\gamma_{\rho]}+\bar{\epsilon
}\frac{i}{\sqrt{2}}e^{-\phi}\partial^{\nu}\phi C_{[\sigma|\nu}\gamma_{\rho
]}-\bar{\epsilon}\frac{i}{\sqrt{2}}e^{-\phi}C_{\sigma\rho}\not \partial
\phi\nonumber\\
&  +\bar{\epsilon}\frac{e^{\phi}}{\sqrt{2}}\gamma_{5}\partial_{\lbrack\sigma
}\mathbf{a}C_{\rho]\mu}\gamma^{\mu}+\bar{\epsilon}e^{2\phi}\gamma_{5}\frac
{i}{2}\nabla_{\lbrack\sigma|}\partial_{\mu}\mathbf{a}\gamma^{\mu}\gamma
_{|\rho]}-\bar{\epsilon}\frac{1}{4}e^{4\phi}\partial_{\mu}\mathbf{a}%
\partial^{\mu}\mathbf{a}\gamma_{\sigma\rho}+\bar{\epsilon}\frac{1}{2}e^{4\phi
}\not \partial \mathbf{a}\partial_{\lbrack\sigma}\mathbf{a}\gamma_{\rho
]}\nonumber\\
&  +\bar{\epsilon}\frac{e^{\phi}}{\sqrt{2}}\gamma_{5}\partial^{\mu}%
\mathbf{a}C_{[\sigma|\mu}\gamma_{|\rho]}-\bar{\epsilon}\frac{e^{\phi}}%
{\sqrt{2}}\gamma_{5}C_{\sigma\rho}\not \partial \mathbf{a}-\bar{\epsilon
}ie^{2\phi}\partial_{\lbrack\sigma}\phi\partial_{\rho]}\mathbf{a}\gamma
_{5}+\bar{\epsilon}\frac{ie^{2\phi}}{2}\not \partial \mathbf{a}\partial
_{\lbrack\sigma}\phi\gamma_{\rho]}\gamma_{5}\nonumber\\
&  -\bar{\epsilon}\frac{e^{\phi}}{\sqrt{2}}\gamma_{5}\not \partial
\mathbf{a}\gamma_{\lbrack\sigma}C_{\rho]\mu}\gamma^{\mu}+\bar{\epsilon}%
\frac{i}{2}e^{2\phi}\gamma_{5}\partial_{\mu}\mathbf{a}\partial^{\mu}\phi
\gamma_{\sigma\rho}-\bar{\epsilon}\frac{i}{2}e^{2\phi}\gamma_{5}%
\not \partial \phi\partial_{\lbrack\sigma}\mathbf{a}\gamma_{\rho]}%
\end{align}

\bigskip

As a matter of fact, in order to evaluate the existence of a Killing spinor in
this theory, one must study both equation $\delta\bar{\chi}=0$ from
(\ref{unmedio-transf}), and (\ref{consistency}) which respectively have the
form%
\begin{equation}
\bar{\epsilon}\Theta=0\quad\text{ and }\quad\bar{\epsilon}\Xi_{\mu\nu}=0\ ,
\label{consist}%
\end{equation}
where $\Theta$ is a $16\times16$ matrix and $\Xi_{\mu\nu}$ are six, $16\times16$
matrices acting on the column arrange $\bar{\epsilon}$ of $16$ components,
which belongs to the tensor product of the vector space of the spinors, times
the vector space of the fundamental representation of $SU\left(  2\right)
\times SU\left(  2\right)  $.

\bigskip

The problem of providing partial classifications of BPS configurations in the
Freedman-Schwarz model has been considered in the literature. As mentioned
above, in \cite{GF} the authors found BPS solutions which are product
spacetimes of the form $AdS_{2}\times \mathbb{R}^{2}$ where the $AdS_{2}$ factor emerges naturally since the gauge fields contribute to the dilaton effective
potential providing an extremum that leads to an effective, two-dimensional,
negative cosmological constant. Such configurations may preserve one-quarter or
one-half of the supersymmetry. Going beyond the product space ansatz, in
\cite{CVcorto,CVlargo} the authors constructed a 1/4 BPS soliton,
which is asymptotically locally flat and it is supported by a single gauge
field (see \cite{Maldacena:2000yy} for further properties of the uplift of the soliton to ten  dimensions). Some non-supersymmetric dyonic solutions where found in \cite{Correa:2003si}, while in \cite{Radu:2002za} the author constructed planar, spherical and hyperbolic solutions of the first order BPS system, both analytically and numerically.

Restricting to the Abelian sector of both gauge fields in a double
dyonic ansatz, in \cite{Klemm} the author constructed planar BPS black holes
and identified a family of singular domain walls as supersymmetric configurations, which were previously
integrated in \cite{Cowdall:1997fn,Singh:1998vf}.
Notwithstanding this, the analysis of \cite{Klemm} provides no new supersymmetric
configurations in the spherically symmetric case.

\bigskip

In the present work we will reconsider this problem. In Section II, we will present a new supersymetric soliton. The solutions is regular, 1/4 BPS and can be obtained from a double Wick rotation of a non-supersymetric configuration found in \cite{Klemm}. In Section III, also in the Abelian sector of the theory we show that there are supersymmetric solutions in the
spherical case. This new 1/4 BPS solutions
describe spacetimes which are singular. These spacetimes are characterized by
two integration constant, and Appendix A is devoted to the explicit presentation of the Killing spinors. Then, in Section IV we introduce the hedgehog
ansatz for a meron gauge field, and construct new solutions of the
$\mathcal{N}=4$ $SU\left(  2\right)  \times SU\left(  2\right)  $ gauged
theory. We combine this magnetic meron ansatz with both a non-Abelian electric field or a second meronic field. Out of the new configurations, we show that only the purely
magnetic, double meron leads to a supersymmetric solution. In Section V we move beyond the supergravity theory, keeping the field content
fixed, but considering more general potentials $V\left(  \phi\right)  $ that
still lead to analytic solutions with interesting thermal properties. The
potentials we consider were already identified as well-behaved potentials
regarding the construction of exact solutions in field theories with similar
matter content (see e.g. \cite{Tarrio:2011de}-\cite{Astefanesei:2021ryn}). For simplicity we will
set the axion field $\mathbf{a}\left(  x\right)  =0$. We provide some conclusions and further comments in Section VI.

\bigskip

\section{New BPS soliton}
The following configuration for the metric, the dilaton and the gauge fields,
provides a solution of $\mathcal{N}=4\ SU(2)\times SU\left(  2\right)  $ gauged
supergravity%
\begin{align}\label{solitonrho}
ds^{2}  &  =\rho dt^{2}-\frac{d\rho^{2}}{g\left(  \rho\right)  }-g\left(
\rho\right)  d\varphi^{2}-\rho dx^{2}\ ,\\
g\left(  r\right)   &  =\frac{\left(  e_{A}^{2}+e_{B}^{2}\right)  }{2}%
\rho-m-\frac{2\left(  Q_{A}^{2}+Q_{B}^{2}\right)  }{\rho}\ ,\\
\phi &  =-\frac{1}{2}\ln\rho\ ,\\
A  &  =\frac{Q_{A}}{\rho}d\varphi\alpha^{3}\text{ and }B=\frac{Q_{B}}{\rho
}d\varphi\beta^{3}\label{gfsoliton}\ ,
\end{align}
where the coordinate $\rho_{0}\leq\rho$ and $\varphi\in\lbrack0,\beta
_{\varphi}]$, with%
\begin{align}
\rho_{0}  &  =\frac{m+\sqrt{4\left(  e_{A}^{2}+e_{B}^{2}\right)  \left(
Q_{A}^{2}+Q_{B}^{2}\right)  +m^{2}}}{\left(  e_{A}^{2}+e_{B}^{2}\right)
}\ ,\\
\beta_{\varphi}  &  =\frac{4\pi}{g^{\prime}\left(  \rho_{0}\right)  }%
=\frac{8\pi\rho_{0}^{2}}{\left(  e_{A}^{2}+e_{B}^{2}\right)  \rho_{0}%
^{2}+4\left(  Q_{A}^{2}+Q_{B}^{2}\right)  ^{2}}\ .
\end{align}
Here $m$ is an integration constants, and we have consistently removed a pure
gauge, second integration constant $\phi_{0}$ that emerges from the
integration of the system. The spacetime \eqref{solitonrho} is regular everywhere and
describes a charged soliton, which asymptotically approaches%
\begin{equation}
ds^{2}=\rho dt^{2}-\frac{2d\rho^{2}}{\left(  e_{A}^{2}+e_{B}^{2}\right)  \rho
}-\frac{\left(  e_{A}^{2}+e_{B}^{2}\right)  }{2}\rho d\varphi^{2}-\rho
dx^{2}\ .
\end{equation}
Notice that the asymptotic geometry acquires an extra conformal Killing vector
which acts as $\rho\rightarrow\lambda \rho$. One can show that the soliton is asymptotically locally flat, since all the components of the Riemann tensor vanish as $\rho$ goes to infinity.

Let us now move to the analysis of the supersymmetry of this solution. As
mentioned above, the consistency and integrability conditions in this theory
reduce to the analysis of the matrices $\Theta$ and $\Xi_{\mu\nu}$ which are
of $16\times16$. For our soliton configuration \eqref{solitonrho}-\eqref{gfsoliton}, the determinant of $\Theta$ reads
\begin{equation}
\det\Theta=\frac{1}{2^{24}\rho^{16}}\left(  4\left(  e_{B}Q_{A}-e_{A}%
Q_{B}\right)  ^{2}+m^{2}\right)  ^{2}\left(  4\left(  e_{B}Q_{A}+e_{A}%
Q_{B}\right)  ^{2}+m^{2}\right)  ^{2}\ .
\end{equation}
Therefore the solution can be supersymmetric only if $m=0$ and%
\begin{equation}
e_{B}Q_{A}\pm e_{A}Q_{B}=0\ .\label{constTheta}%
\end{equation}
Setting $m=0$ in $\Xi_{\mu\nu}$ leads to the following determinant%
\begin{equation}
\det\left(  \Xi_{\rho\varphi}\right)  =\left(  \frac{1}{2\rho}\right)
^{32}\left(  e_{B}Q_{A}-e_{A}Q_{B}\right)  ^{8}\left(  e_{B}Q_{A}+e_{A}%
Q_{B}\right)  ^{8}\ ,
\end{equation}
which actually vanishes identically given the BPS constraint coming from
(\ref{constTheta}). All the remaining determinants identically vanish, even before using (\ref{constTheta}). In summary, this implies that the configuration \eqref{solitonrho}-\eqref{gfsoliton} with
$m=0$ and $e_{B}Q_{A}=\mp e_{A}Q_{B}$ is supersymmetric. This solution turns
out to be 1/4 BPS, and the explicit expression for the Killing spinors is
presented in Appendix A. The supersymmetric solution with $m=0$ takes a
particularly simple form after performing the change of coordinates%
\begin{equation}
\rho=\frac{2Q_{A}}{e_{A}}\cosh l\ ,
\end{equation}
with $0\leq l<+\infty$, since it reduces to%
\begin{equation}
ds_{\text{BPS-Soliton}}^{2}=\frac{Q_{A}}{e_{A}}\cosh l\ \left[  2dt^{2}%
-\frac{4dl^{2}}{e_{A}^{2}+e_{B}^{2}}-\left(  e_{A}^{2}+e_{B}^{2}\right)
\tanh^{2}ld\varphi^{2}-2dx^{2}\right]  \ .
\end{equation}

\bigskip

Our new solitonic solution can also be obtained from a double analytic
continuation of the planar solution  found in \cite{Klemm}. Such
spacetime is characterized by an integration constant $\tilde{m}$, that maps
to our $m$ after the continuation. Nevertheless, the planar solution found in
\cite{Klemm} with $\tilde{m}=0$ is a naked singularity. Notwithstanding this fact,
the double analytic continuation, followed by a compactification of the
coordinate $\varphi$ with an appropriate range, leads to a completely regular
soliton spacetime. The same effect has been recently seen to work for
$\mathcal{N}=2$ gauged supergravity in $D=4$ and $D=5$ in \cite{AnabalonRoss}. The new 1/2 BPS solitons discovered in
such reference are connected, via a double analytic continuation, to the
plannar Reissner-Norstrom-AdS spacetime for a value of the integration
constant that would lead to a naked singularity in the latter. Remarkably, the
double Wick rotation leads to a smooth spacetime with unbroken supersymmetries\footnote{Double analytic continuations may also give rise to supersymmetric wormholes, when the seed spacetime is Taub-NUT-AdS in the hyperbolic foliation \cite{AdWO}.}
\cite{AnabalonRoss}, as we have also reported here for $\mathcal{N}=4\ SU(2)\times
SU\left(  2\right)  $ gauged supergravity.

\section{New Abelian BPS configuration}

In this section we reconsider the problem of spherically symmetric,
supersymmetric solutions of the Freedman-Schwarz model in the Abelian sector.
In order to compare with reference \cite{Klemm}, we consider the following ansatz
for the metric\footnote{Notice that in \cite{Klemm} there is an extra $\gamma$
factor in front of the sphere. Such factor can be gauged-away in the solution
by an appropriate redefinition of the radial and time coordinates, including a shift in the integration constant $\phi_{0}$ of the dilaton.}%
\begin{equation}
ds^{2}=f\left(  \rho\right)  dt^{2}-\frac{d\rho^{2}}{f\left(  \rho\right)
}-\rho\left(  d\theta^{2}+\sin^{2}\theta d\varphi^{2}\right)  \ .
\label{metricrho}%
\end{equation}
Going to the areal radial gauge is trivially achieved by introducing the
coordinate $r$ such that $\rho=r^{2}$. We consider the following Abelian,
dyonic ansatz for the gauge fields%
\begin{align}
A  &  =\left(  \frac{Q_{A}e^{2\phi_{0}}}{\rho}dt-H_{A}\cos\theta
d\varphi\right)  \alpha^{3}\ ,\\
B  &  =\left(  \frac{Q_{B}e^{2\phi_{0}}}{\rho}dt-H_{B}\cos\theta
d\varphi\right)  \beta^{3}\ ,
\end{align}
where $\phi_{0}$ is
\begin{equation}
\phi_{0}=\frac{\ln\left(  2\left(  H_{A}^{2}+H_{B}^{2}\right)  \right)  }%
{2}\ ,
\end{equation}
while the dilaton reads%
\begin{equation}
\phi=\phi_{0}-\frac{1}{2}\ln\rho\ .
\end{equation}
The field equations are solved by%
\begin{equation}
f\left(  \rho\right)  =\left(  1+\left(  g_{A}^{2}+g_{B}^{2}\right)  \left(
H_{A}^{2}+H_{B}^{2}\right)  \right)  \rho-m+\frac{4\left(  Q_{A}^{2}+Q_{B}%
^{2}\right)  \left(  H_{A}^{2}+H_{B}^{2}\right)  }{\rho}\ , \label{fderho1}%
\end{equation}
with the constraint%
\begin{equation}
H_{A}Q_{A}+H_{B}Q_{B}=0\ ,
\end{equation}
which comes from the field equation of the vanishing axion.

Depending on the relation between the integration constant $m$ and the
remaining charges, the spacetime (\ref{metricrho}) with the function
(\ref{fderho1}) may describe a black hole, which can be extremal. The black
hole asymptotically matches the metric%
\begin{equation}
ds^{2}=\left(  1+\left(  g_{A}^{2}+g_{B}^{2}\right)  \left(  H_{A}^{2}%
+H_{B}^{2}\right)  \right)  r^{2}dt^{2}-\frac{dr^{2}}{\left(  1+\left(
g_{A}^{2}+g_{B}^{2}\right)  \left(  H_{A}^{2}+H_{B}^{2}\right)  \right)
}-r^{2}\left(  d\theta^{2}+\sin^{2}\theta d\varphi^{2}\right)  \ ,
\end{equation}
which is
asymptotically, locally flat since $R_{\ \ \alpha\beta}^{\mu\nu}$ goes to zero
as $r\rightarrow\infty$.

\bigskip

First, we are interested in revisiting the supersymmetry of this family of
solutions by analyzing the consistency equations $\bar{\epsilon}\Theta=0$ and
$\bar{\epsilon}\Xi_{\mu\nu}=0$ in (\ref{consist}). As in reference \cite{Klemm}, we
obtain that the $16\times 16$ matrix $\Theta$ is block diagonal, and has the form%
\begin{equation}
\Theta=\left(
\begin{array}
[c]{cc}%
\Theta_{+} & 0\\
0 & \Theta_{-}%
\end{array}
\right)  \ ,
\end{equation}
where $\Theta_{\pm}$ are $8\times8$ matrices, with determinants%
\begin{equation}
\det\left(  \Theta_{\pm}\right)  =r^{-8}\left(  K_{\pm}^{0}+rK_{\pm}^{1}%
+r^{2}K_{\pm}^{2}\right)  \ .
\end{equation}
The functions $K_{\pm}^{0,1,2}$ depend only on the integration constants.
Requiring a nontrivial solution of $\bar{\epsilon}\Theta=0$ for $\bar
{\epsilon}$, obviously implies $\det\Theta_{+}=0$ or $\det\Theta_{-}=0$, which
leads to%
\begin{align}
\left(  m\pm4\left(  H_{B}Q_{A}-H_{A}Q_{B}\right)  \right)  ^{2}-16\left(
H_{A}^{2}+H_{B}^{2}\right)  ^{4}\left(  Q_{A}e_{B}\mp Q_{B}e_{A}\right)  ^{2}
&  =0\label{constrainttheta1}\\
H_{A}e_{A}\pm H_{B}e_{B}  &  =0 \label{constrainttheta2}%
\end{align}

On the other hand, the six, $16\times16$ matrices $\Xi_{\mu\nu}$ that can be read from
(\ref{consist}), coming from the consistency condition of the variation of the
Rarita-Schwinger field, also acquire a block-diagonal structure with
non-trivial $8\times8$ blocks. The only components of $\Xi_{\mu\nu}$ in the consistency condition \eqref{consist} with non-vanishing determinants, lead to the following expressions%
\footnotesize
\begin{align}
0  &  =\det\left(  \Xi_{tr}\right)  =\frac{1}{\left(  8r^{2}\right)  ^{16}%
}\left(  -16\left(  Q_{A}e_{B}-Q_{B}e_{A}\right)  ^{2}\left(  H_{A}^{2}%
+H_{B}^{2}\right)  ^{2}-16\left(  Q_{A}^{2}+Q_{B}^{2}\right)  \left(
H_{A}^{2}+H_{B}^{2}\right)  +m^{2}\right)  ^{4}\label{constraintRS2}\\
&  \qquad\qquad\qquad\qquad\qquad\times\left(  -16\left(  Q_{A}e_{B}%
+Q_{B}e_{A}\right)  ^{2}\left(  H_{A}^{2}+H_{B}^{2}\right)  ^{2}-16\left(
Q_{A}^{2}+Q_{B}^{2}\right)  \left(  H_{A}^{2}+H_{B}^{2}\right)  +m^{2}\right)
^{4}\label{constraintRS1}\\
0  &  =\det\left(  \Xi_{\theta\phi}\right)  =\left(  \frac{\sin\theta}%
{4}\right)  ^{16}\left(  g_{A}H_{A}+g_{B}H_{B}\right)  ^{8}\left(  g_{A}%
H_{A}-g_{B}H_{B}\right)  ^{8}%
\end{align}
\normalsize
For $\det\Theta_{\pm}=0$ and $\det\left(  \Xi_{tr}\right)  =0$, we have obtained the same equations than in reference \cite{Klemm}. Nevertheless, our
expression for $\det\left(  \Xi_{\theta\phi}\right)  $ differs from the one
reported in \cite{Klemm}, and in our case it can vanish for a suitable relation between the magnetic charges, which leads to a novel supersymmetric solution. 
Implementing
all the supersymmetric constraints (\ref{constrainttheta1}),
(\ref{constrainttheta2}), (\ref{constraintRS2}) and (\ref{constraintRS1}), the
metric function $f\left(  \rho\right)  $ reduces to 
\begin{equation}
f_{BPS}=\frac{((e_{A}^2+e_{B}^2)^2H_{A}^2+e_{B}^2)}{e_{B}^2}\rho-m_{BPS}+\frac{4Q_{A}^2(e_A^2+e_B^2)^2H_{A}^2}{e_{A}^2e_{B}^2\rho}  \ ,\label{fBPS}
\end{equation}
where%
\begin{equation}
m_{BPS}=\frac{4H_{A}Q_{A}\left(  e_{A}^{2}+e_{B}^{2}\right)  }{e_{A}e_{B}}\ .
\end{equation}
Notice that the expression for $m_{BPS}$ does not depend on the sign choice
made in (\ref{constrainttheta1})-(\ref{constrainttheta2}), therefore without
loosing generality one can restrict to one of the two signs. One can see that
$f_{BPS}\left(  \rho\right)  $ in (\ref{fBPS}) does not vanish, and since the
spacetime (\ref{metricrho}) has a singularity at the origin $\rho=0$, this
solution represents a BPS naked singularity. It is well-known that singular
spacetimes can indeed fulfill BPS conditions, as it is the case of the
Reissner-Norstrom-AdS solution in $\mathcal{N}=2$ $U\left(  1\right)  $ gauged
supergravity, with the mass equal to the charge \cite{Romans}. After a simple counting one sees that our BPS solution
depends on two arbitrary integration constants.

The BPS background obtained with $Q_{A}=0$, which implies $m_{BPS}=0$, does
not lead to an enhancement of the supersymmetries, and it is actually a
singular spacetime, which has a divergence at the origin that is milder than
the singularity at the origin of a Schwarzschild black hole, since in this
case setting $H_{A}=H_{B}=m=0$ in (\ref{fderho1}) leads to a Kretchmann scalar
that diverges at the origin as $r^{-4}$, where $r$ is the areal radial coordinate.

\bigskip

Notice that, as explained in \cite{Klemm} for planar black holes, the extremal
configurations are 1/4 BPS, while the background obtained by setting
to zero all the integration constants, which is also the metric approached by
the black holes at infinity, do acquire some extra supersymmetry leading to
1/2 BPS solutions. As explained here, the situation for the
spherically symmetric case is different and both, the metric deformed by the
non-vanishing value of the charges, as well as the asymptotic metric, have
the same amount of unbroken supersymmetry, namely they are 1/4 BPS

\bigskip

The solutions we have identified as BPS, preserves one-quarter of the
supersymmetry and one can pursue the explicit integration of the Killing spinors, which
leads to the structure%
\begin{align}
    \bar{\epsilon}_{i}=\Psi_{1}(\rho)D_{i}+\Psi_{2}(\rho)E_{i} \ .%
\end{align}
The details are given in Appendix A.

The metric (\ref{metricrho}) with the function (\ref{fderho1}) has four isometries, which
close in the algebra $\mathbb{R}\times so\left(  3\right)  $. These are generated by
the Killing vectors $\partial_{t}$ plus the usual three Killing vectors of the
round sphere spanned in spherical coordinates. We have explicitly checked that
the bilinears%
\begin{equation}
K^{\mu}=\bar{\epsilon}_{i}\gamma^{\mu}\epsilon_{i}\text{ no sum in }i\ ,
\end{equation}
do indeed lead to the isometries of the spacetime, for each of the four
independent Killing spinors. 

\section{Charged merons and double meron in gauged supergravity}

Hereafter we work with the following gauge for the metric%
\begin{equation}
ds^{2}=N\left(  r\right)  f\left(  r\right)  dt^{2}-\frac{dr^{2}}{f\left(
r\right)  }-r^{2}\left(  d\theta^{2}+\sin^{2}\theta d\varphi^{2}\right)  \ ,
\label{metricNyf}%
\end{equation}
while the dilaton still depends only on the radial coordinate, namely
$\phi=\phi\left(  r\right)  $.

\subsection{Charged meron}

Let us consider the following ansatz for the gauge fields
\begin{align}
A  &  =\xi U_{\left(  \alpha\right)  }^{-1}dU_{\left(  \alpha\right)
}\ ,\text{ }U_{\left(  \alpha\right)  }\left(  x^{\mu}\right)  \in SU\left(  2\right)  \ ,
\label{meron}\\
B  &  =\frac{Q_{B}}{r^{2}}dt\ \beta_{3}\ . \label{electric}%
\end{align}
This ansatz corresponds to a superposition of an electric Abelian SU(2) gauge field and a
meron configuration \cite{deAlfaro:1976qet}, the
latter being proportional to the Maurer-Cartan left-invariant form of
$su\left(  2\right)  $. It is interesting to notice that meron gauge fields also lead to black holes supported by non-Abelian gauge fields \cite{Canfora:2012ap,Canfora:2018ppu}, on which the Jackiw, Rebbi, Hasenfratz, ’t Hooft mechanism of spin from isospin is present \cite{Jackiw:1976xx}-\cite{Hasenfratz:1976gr}. We further specialize the expressions for the meron to the hedgehog
ansatz, in terms of a group valued function $U$ given by%
\begin{align}
U_{\left(  \alpha\right)  }^{\pm1}  &  =\boldsymbol{1}\cos\Upsilon_{\left(
\alpha\right)  }\left(  r\right)  \pm\sin\Upsilon_{\left(  \alpha\right)
}\left(  r\right)  \hat{x}^{i}\alpha_{i}\ ,\\
\hat{x}^{1}  &  =\sin\theta\cos\varphi\ ,\quad\hat{x}^{2}=\sin\theta
\sin\varphi\ ,\quad\hat{x}^{3}=\cos\theta\ .
\end{align}
Here we are using the generator of $su\left(  2\right)  \times su\left(
2\right)  $ given (\ref{losalphas}) and (\ref{losbetas}). Setting
$\Upsilon\left(  r\right)  =\pi/2$ we can substantially simplify the
equations; the new solutions we find below, belong to such sector. The
Yang-Mills equations as well as the equation for the dilaton are fulfilled when
the constant $\xi$ is fixed as%
\begin{equation}
\xi=-\frac{1}{e_{A}}\ ,
\end{equation}
and the dilaton is given by%
\begin{equation}
\phi=-\ln\left(  \frac{e_{A}r}{\sqrt{2}}\right)  \ ,
\end{equation}
therefore the gauge field $A$ and the dilaton are devoid of integration
constants. Consequently, the field strength associated to the gauge field $A$
reads%
\begin{equation}
A_{\left[  2\right]  }=\frac{1}{2}A_{\mu\nu}dx^{\mu}\wedge dx^{\nu}=\frac
{1}{2e_{A}}\left(  1-\frac{1}{2e_{A}}\right)  \left[  U_{\left(
\alpha\right)  }^{-1}\partial_{\mu}U_{\left(  \alpha\right)  },U_{\left(
\alpha\right)  }^{-1}\partial_{\nu}U_{\left(  \alpha\right)  }\right]
dx^{\mu}\wedge dx^{\nu}\ ,
\end{equation}
which implies that if the meron ansatz (\ref{meron}) turns out to be aligned
with a single generator, its field strength would vanish. With these
definitions, the gauge field $A_{\mu}$ has the following explicit form%
\footnotesize
\begin{equation}
A=-e_{A}^{-1}\left[  \left(  -\sin\varphi d\theta-\cos\theta\sin\theta
\cos\varphi d\varphi\right)  \alpha_{1}+\left(  \cos\varphi d\theta-\cos
\theta\sin\theta\sin\varphi d\varphi\right)  \alpha_{2}+\sin^{2}\theta
d\varphi\alpha_{3}\right]  \ . \label{Ameronexplicit}%
\end{equation}
\normalsize

Some remarks on the nature of this ansatz are now in order. The configuration \eqref{Ameronexplicit} is related to an Abelian one by the group element\footnote{We thank Andrés Anabalón for bringing this transformation to our attention.} $g=e^{\frac{1}{2}\theta\alpha_2}e^{\frac{1}{2}\phi\alpha_3}$, such that locally $A_{\text{monopole}}=gA_{\text{meron}}g^{-1}-\frac{2}{e_{A}}gdg^{-1}$. Given the fact that $g$ does not go to the identity at infinity and it it not continuous as $r\rightarrow0$, the two configurations $A_{\text{monopole}}$ and $A_{\text{meron}}$ are not gauge equivalent, since two gauge fields are gauge equivalent when they are related by a \textit{smooth gauge transformation} which approaches an element of the center of the gauge group at spatial infinity. The discontinuity at the origin can be seen clearly by considering that
\begin{align}
g(x=0,y=0,z=0^+)&=\begin{pmatrix}
i & 0 \\
0 & -i
\end{pmatrix} \ ,
\end{align}
while
\begin{align}
g(x=0,y=0,z=0^-)&=\begin{pmatrix}
0 & -i \\
-i & 0
\end{pmatrix}\ .
\end{align} 
Such discontinuity will be a feature of any transformation relating both configurations, which confirms that they belong to different gauge equivalence classes. For a more detailed discussion, see Section V of  \cite{Canfora:2012ap}. Furthermore the transformation that relates \eqref{Ameronexplicit} and $A_{\text{monopole}}$ has a non-trivial winding, and therefore it is topologically non-trivial. This can be seen as follow. If the two configurations were globally gauge equivalent, it should be fulfilled that $F_{\text{monopole}}=gF_{\text{meron}}g^{-1}$. Nevertheless, the two configurations are related by
\begin{equation}\label{casigauge}
F_{\text{monopole}}=gF_{\text{meron}}g^{-1}+d\left(gdg^{-1}\right)-dg\wedge dg^{-1}\ .
\end{equation} For globally defined gauge transformations the last two terms in this equation cancel each other, nevertheless, this is not the case for the transformation generated by the group element $g=e^{\frac{1}{2}\theta\alpha_2}e^{\frac{1}{2}\phi\alpha_3}$, which in turn can be seen integrating equation \eqref{casigauge} on the two dimensional surface $0\leq r\leq 1$ and $0\leq\phi<2\pi$ with $\theta=\theta_0$ and $t=\text{constant}$. Such integration leads to
\begin{equation}
\int_{\text{disk}}F_{\text{monopole}}=\int_{\text{disk}}gF_{\text{meron}}g^{-1}+\int_{\partial\text{disk}=S^1}gdg^{-1}\ ,
\end{equation}
and the last term is non-vanishing and given by
\begin{equation}
w\left(gdg^{-1}\right)=\int_{\partial\text{disk}=S^1}gdg^{-1}=\pi\sin(\theta_0)\alpha_1-\pi\cos(\theta_0)\alpha_3\ .
\end{equation}
This argument reinforces the fact that the meron configuration \eqref{Ameronexplicit} cannot be identified as physically equivalent with the monopole configuration.

\bigskip

The metric of the spacetime is in the family of spherically symmetric
solutions (\ref{metricNyf}) and it is explicitly given by
\begin{equation}
ds^{2}=r^{2}\left(  \frac{1}{2}+\frac{e_{B}^{2}}{4e_{A}^{2}}+\frac{e_{A}%
^{2}Q_{B}^{2}}{r^{4}}-\frac{\mu}{r^{2}}\right)  dt^{2}-\frac{dr^{2}}{\left(
\frac{1}{2}+\frac{e_{B}^{2}}{4e_{A}^{2}}+\frac{e_{A}^{2}Q_{B}^{2}}{r^{4}%
}-\frac{\mu}{r^{2}}\right)  }-r^{2}\left(  d\theta^{2}+\sin^{2}\theta
d\varphi^{2}\right)  \ , \label{chargedmeron}%
\end{equation}
where $\mu$ is an integration constant. The metric (\ref{chargedmeron})
represents a black hole which can have an event horizon at $r=r_{+}$, as well as an
inner Cauchy horizon located at $r=r_{-}$, while the integration constant $\mu$ is related to the
mass of the black hole. The location of the horizons are%
\begin{equation}
r_{\pm}=\sqrt{\frac{2e_{A}^{2}}{2e_{A}^{2}+e_{B}^{2}}\left(  \mu\pm\sqrt
{\mu^{2}-\left(  2e_{A}^{2}+e_{B}^{2}\right)  Q_{B}^{2}}\right)  }\ .
\end{equation}
Black hole solutions are present when
\begin{equation}
\mu^{2}\geq\left(  2e_{A}^{2}+e_{B}^{2}\right)  Q_{B}^{2}\ ,
\end{equation}
and the spacetime becomes an extremal black hole when the bound is
saturated. Asymptotically, the spacetime is locally flat and the scalar field $\phi\sim-\ln\left(  r\right)  $, reaches the absolute maximum of
the potential (\ref{Vsugra}).

The Hawking temperature in terms of $r_{+}$ reads%
\begin{equation}
T=\frac{2e_{A}^{2}+e_{B}^{2}}{8\pi e_{A}^{2}}-\frac{e_{A}^{2}Q_{B}^{2}}{2\pi
r_{+}^{4}}\ .
\end{equation}
Notice that in the absence of the non-Abelian electric charge, namely for
$Q_{B}=0$, the temperature reduces to a constant. The integration constant
$\mu$ does not appear in the temperature when $Q_{B}=0$. If one pushes forward
the interpretation of $\mu$ as the energy content of the spacetime, the family
of black holes with $Q_{B}=0$ would lead to a divergence in the heat capacity
$C\sim\frac{\partial\mu\left(  T,Q\right)  }{\partial T}$, which may be
interpreted as a sign of criticality. Therefore the non-Abelian gauge field is needed in order to properly define the thermodynamics of these configurations. 

\bigskip

Now, the analysis of the supersymmetry of this backgrounds is in order. In the presence of the meron ansatz (\ref{meron}), the analysis of
supersymmetry is again dictated by the structure of the matrices $\Theta$ and
$\Xi_{\mu\nu}$, defined in the previous sections. Actually in this case, it is
enough to analyze the range of the matrix $\Xi_{\theta\phi}$. In fact
\begin{equation}
0=\det\left(  \Xi_{\theta\phi}\right)  =\left(  \frac{\sin\theta}{4}\right)
^{16}\ .
\end{equation}
This shows that the electric-meronic configuration we have constructed in this
section, cannot preserve any supersymmetry.

\bigskip

In the next section, we move to the double meron configuration.

\bigskip

\subsection{Double meron}

Now, we introduce a meron ansatz in each of the $su\left(  2\right)  $ factors, namely
\footnotesize
\begin{align}
A  &  =\xi_{A}\left[  \left(  -\sin\varphi d\theta-\cos\theta\sin\theta
\cos\varphi d\varphi\right)  \alpha_{1}+\left(  \cos\varphi d\theta-\cos
\theta\sin\theta\sin\varphi d\varphi\right)  \alpha_{2}+\sin^{2}\theta
d\varphi\alpha_{3}\right]  \ ,\\
B  &  =\xi_{B}\left[  \left(  -\sin\varphi d\theta-\cos\theta\sin\theta
\cos\varphi d\varphi\right)  \beta_{1}+\left(  \cos\varphi d\theta-\cos
\theta\sin\theta\sin\varphi d\varphi\right)  \beta_{2}+\sin^{2}\theta
d\varphi\beta_{3}\right]  \ .
\end{align}
\normalsize
Again, Yang-Mills and the dilaton equations are satisfied if%
\begin{equation}
\xi_{A}=-\frac{1}{e_{A}}\text{\ },\ \xi_{B}=-\frac{1}{e_{B}}\ ,
\end{equation}
and the dilaton field takes the form%
\begin{equation}
\phi\left(  r\right)  =-\ln\left(  \frac{e_{B}e_{A}}{\sqrt{2}\sqrt{e_{A}%
^{2}+e_{B}^{2}}}r\right)  \ .
\end{equation}
Notice that as it is the case in the electrically charged meron, the gauge
fields and the dilaton are completely fixed in terms of the couplings of the
theory and are devoid of any integration constant. In this case the spacetime
metric reduces to
\begin{equation}
ds^{2}=r^{2}\left(  \tilde{\Lambda}^2-\frac{\mu}{r^{2}}\right)  dt^{2}%
-\frac{dr^{2}}{\tilde{\Lambda}^2-\frac{\mu}{r^{2}}}-r^{2}\left(  d\theta
^{2}+\sin^{2}\theta d\varphi^{2}\right)  \ , \label{bhdouclemeron}%
\end{equation}
where $\mu$ is an integration constant and
\begin{equation}
\tilde{\Lambda}^2:=\frac{e_{A}^{4}+3e_{A}^{2}e_{B}^{2}+e_{B}^{4}}{4e_{A}%
^{2}e_{B}^{2}}>0\ .
\end{equation}
When $\mu>0$, this metric describes a black hole with an event horizon located
at $r=r_{+}=\left(  \mu/\tilde{\Lambda}^2\right)  ^{1/2}$. It is worth emphasize that in \eqref{bhdouclemeron} the $1/r^4$ term is absent, in contrast with \eqref{chargedmeron}. This implies that the electric and magnetic parts enter on different footing in these configurations. One way to understand this lack of democracy between electric and magnetic fields is the presence of the dilaton, which changes the electromagnetic duality properties of the theory. 

Now, we move to the analysis of the supersymmetry of the double meron family.
The non-vanishing integrability conditions for the Killing spinor,
$\bar{\epsilon}\Theta=0$ and $\bar{\epsilon}\Xi_{\mu\nu}=0$ in (\ref{consist}%
) read%
\begin{align}
\det\Xi_{tr}  &  =\frac{\mu^{16}}{r^{48}}\ ,\nonumber\\
\det\Theta &  =\frac{\left(  r^{2}-\mu\right)  ^{2}\left(  r^{2}+\mu\right)
^{2}\mu^{4}}{256r^{32}}\ .
\end{align}
Therefore, the background of the black hole (\ref{bhdouclemeron}) with $\mu=0$
preserves some supersymmetry. This background metric is actually one-quarter
BPS, and the metric, which is also recovered as the asymptotic geometry of the
black holes (\ref{bhdouclemeron}) reduces to%
\begin{equation}
ds^{2}=\tilde{\Lambda}^2r^{2}dt^{2}-\tilde{\Lambda}^{-2}dr^{2}-r^{2}\left(
d\theta^{2}+\sin^{2}\theta d\varphi^{2}\right)  \ .
\end{equation}
This background possesses only the obvious Killing vectors as
isometries,\ namely the time translation and the $SO\left(  3\right)  $
Killing vectors of the sphere at the $r=$constant surfaces of the spacelike
surfaces at $t=$constant. Nevertheless, it is worth mentioning that this
background has an extra conformal Killing vector given by%
\begin{equation}
l=r\partial_{r}\ .
\end{equation}

\bigskip

The temperature of the black hole (\ref{bhdouclemeron}) has the intriguing
property of being independent of $r_{+}$, and it is
given by%
\begin{equation}\label{temperaturedoublemeron}
T=\frac{\tilde{\Lambda}^2}{2\pi}\ .
\end{equation}
Wald's formula for the entropy yields%
\begin{equation}
S=\frac{A}{4G}=4\pi^{2}r_{+}^{2}\ ,
\end{equation}
since in the normalization of the Einstein term in (\ref{Lsugrafull}) we have
chosen $G=(4\pi)^{-1}$. First law%
\begin{equation}
dM=TdS\ ,
\end{equation}
provides the following value for the mass of the black hole%
\begin{equation}
M=2\pi\tilde{\Lambda}^2r_{+}^{2}\ .
\end{equation}
Since the temperature of the black hole does not depend on its radius, we have
that the heat capacity%
\begin{equation}
C=\frac{dM}{dT}\ ,
\end{equation}
diverges, signaling the presence of a critical behavior. The free energy vanishes identically.

\bigskip

The expression for the temperature in \eqref{temperaturedoublemeron} can be considered as evidence of the fact
that the black hole we are currently analyzing is a particular case of a more
general solution in the double meron sector. It would be interesting to see
whether one can design a deformation of our ansatz that would allow to turn on
the axion field $\bold{a}\left(  x\right)  $ in a simple enough manner as to
construct new exact and possibly BPS solutions.

\bigskip

In the next sections, we show that there are families of potentials that go beyond
the supergravity potential (\ref{Vsugra}) and that allow for the construction
of exact hairy black holes. As a matter of fact, we adapt the normalization
and conventions in the action principle for each case, in order to simplify
the presentation and analysis of the exact solutions.

\bigskip

\section{Hairy black holes beyond supergravity}

Since in the previous sections we have seen that the meron ansatz was fruitful in the construction of exact solutions in gauged supergravity, in what follows we will construct exact hairy black holes in the
Einstein-Yang-Mills dilaton theory. Exact solutions in field theories with
similar matter content have already been considered in the literature, see e.g. \cite{Tarrio:2011de}-\cite{Astefanesei:2021ryn}). Here we focus on the
meron sector of the theory, with a single $su\left(  2\right)  $ gauge field.
We will introduce a suitable choice of the coefficient in the dilatonic coupling as well as a
self-interaction for the dilaton, which allow for the construction of
solutions in a closed form. The theories we consider in what follows can be
seen as a truncation of the gauged supergravity of the previous sections,
supplemented by a deformation of the theory that explicitly breaks the supersymmetry.

\subsection{Exponential potential: topologically Lifshitz black holes}

For the first family of potentials it is convenient to introduce the following
suitable parametrization of the action%
\begin{equation}
I\left[  g_{\mu\nu},A,\phi\right]  =\frac{1}{16\pi G}\int d^{4}x\sqrt
{-g}\left(  R-\frac{1}{2e^{2}}e^{-\frac{2}{\sqrt{z-1}}\phi}F_{\mu\nu}%
^{i}F^{i\mu\nu}-\frac{1}{2}\left(  \partial\phi\right)  ^{2}-V_{1}\left(
\phi\right)  \right)  \label{actionVgen1}%
\end{equation}
where the potential is given by%
\begin{equation}
V_{1}\left(  \phi\right)  =\xi e^{\sqrt{z-1}\phi}+\frac{2\left(  z-1\right)
}{\eta^{2}\left(  z-2\right)  }e^{\frac{\phi}{\sqrt{z-1}}}-\frac{\left(
z+2\right)  \left(  z+1\right)  }{l^{2}}\ , \label{potencial1}%
\end{equation}
with
\begin{equation}
\eta:=\left(  \frac{l^{2}}{4e^{2}\left(  z+2\right)  \left(  z-1\right)
}\right)  ^{1/4}\ . \label{eta1}%
\end{equation}
Here the field strength off the $SU(2)$ gauge field $A=A^i_{\mu}dx^{\mu}t_i$ reads
\begin{equation}
F=dA+A\wedge A
\end{equation}
The range $z>1$ will allow us to obtain a topologically Lifshitz asymptotic
behavior. For $z>2$ and $\xi\geq0$, the potential (\ref{potencial1}) is
clearly bounded from below, and it is characterized by three independent
constants, $\xi,$ $l$ and $z$. The gauge coupling $e$ also appears here. In
these cases, the potential takes its minimum value (which is negative), when
$\phi\rightarrow-\infty$. Notice that the potential (\ref{potencial1}) cannot
be continuously connected with the supergravity potential in equation (\ref{Vsugra}).

For this model, the field equations are solved by the following scalar field%
\begin{equation}
\phi\left(  r\right)  =-2\sqrt{z-1}\ln\left(  \frac{r}{\eta}\right)  \ ,
\end{equation}
which approaches the minimum of the potential in the asymptotic region
$r\rightarrow\infty$ since the metric of the spacetime in this case reads%
\begin{equation}
ds^{2}=-\frac{r^{2z}}{l^{2z-2}}g\left(  r\right)  dt^{2}+\frac{dr^{2}}%
{r^{2}g\left(  r\right)  }+r^{2}d\Omega^{2} \label{lif1}%
\end{equation}
with%
\begin{equation}
g\left(  r\right)  =\frac{1}{l^{2}}-\frac{\mu}{r^{z+2}}-\frac{1}{\left(
z-2\right)  z}\frac{1}{r^{2}}+\frac{\eta^{2\left(  z-1\right)  }\xi}{2\left(
z-4\right)  }\frac{1}{r^{2\left(  z-1\right)  }}\ . \label{gparalif1}%
\end{equation}
While the gauge field as in the previous sections reads%
\footnotesize
\begin{equation}
A=\frac{1}{2}\left[  \left(  -\sin\varphi d\theta-\cos\theta\sin\theta
\cos\varphi d\varphi\right)  \alpha_{1}+\left(  \cos\varphi d\theta-\cos
\theta\sin\theta\sin\varphi d\varphi\right)  \alpha_{2}+\sin^{2}\theta
d\varphi\alpha_{3}\right]  \ . \label{meronVs}%
\end{equation}
\normalsize
Here $\mu$ is an integration constant that will determine the energy content
of the spacetime, namely its mass. It is important to stress that the
spacetime configuration (\ref{lif1})-(\ref{gparalif1}) is characterized by a
single integration constant, since all the remaining variables, $z$, $\xi$,
$\eta$, $l$, are determined by the self-interaction (\ref{potencial1}) and by
the definition (\ref{eta1}). Therefore, the black holes have a secondary hair.

Asymptotically, the spacetime defined by (\ref{lif1})-(\ref{gparalif1})
approaches%
\begin{equation}
ds_{\text{asymp}}^{2}=-\frac{r^{2z}}{l^{2z}}dt^{2}+\frac{l^{2}dr^{2}}{r^{2}%
}+r^{2}d\Omega^{2}\ , \label{asymp1}%
\end{equation}
which is the topological extension of a Lifshitz spacetime \cite{Mann:2009yx}. Notice that, as
$r\rightarrow\infty$ the constant $r$ surfaces have an induced light cone
structure which is non-relativistic in such limit, since for $r=r_{c}$ we have%
\begin{equation}
ds_{\text{induced}}^{2}=-\frac{r_{c}^{2z}}{l^{2z}}dt^{2}+r_{c}^{2}d\Omega
^{2}\ ,
\end{equation}
and therefore a massless particle moving on the sphere $S^{2}$, along a
trajectory $\theta=\theta\left(  \lambda\right)  $ and $\phi=\phi\left(
\lambda\right)  $, will fulfil%
\begin{equation}
\frac{d\omega\left(  \lambda\right)  }{dt}=l^{2z}r_{c}^{2z-2}\text{ ,}%
\end{equation}
which goes to infinity as $r_{c}\rightarrow\infty$ when $z>1$. Here we have
defined $d\omega\left(  \lambda\right)  =\sqrt{\left(  \frac{d\theta}%
{d\lambda}\right)  ^{2}+\sin^{2}\theta\left(  \frac{d\phi}{d\lambda}\right)
^{2}}d\lambda$. Lifshitz spacetimes are defined when the sphere in
(\ref{asymp1}) is replaced by flat space. If we use Cartesian coordinates
$\vec{x}$ for such flat space, the corresponding spacetime has the following
anisotropic scaling symmetry $r\rightarrow\chi r$, $\vec{x}\rightarrow
\chi^{-1}\vec{x}$ and $t\rightarrow\chi^{-z}t$, a symmetry that emerges for
non-relativistic systems near criticality \cite{Kachru:2008yh}.
Topologically Lifshitz spacetimes break such scaling symmetry, nevertheless
the non-relativistic interpretation of a potential dual theory living on the
boundary of the spacetime remains, due to the previous argument \cite{Mann:2009yx}. A massless particle travelling radially
takes a finite time to go from a point in the bulk to infinity, and the causal
asymptotic behavior of (\ref{lif1})-(\ref{gparalif1}) is that of AdS,
therefore the spacetime has a timelike boundary.

\bigskip

The spacetime defined by (\ref{lif1}) with (\ref{gparalif1}), has a curvature singularity at
$r=0$, which may be covered by an event and a Cauchy horizon, depending on the
details of the potential as well as on the value of the integration constant
$\mu$.

\bigskip

Since we do not have a supergravity embedding of the potential
(\ref{potencial1}), which may allow us to prove the positivity of the energy,
the cases with potentials bounded from below will be particulary relevant.
This is ensured $z>2$ and $\xi\geq0$.

\bigskip

- For $2<z\leq4$, there is a minimum negative value of $\mu=\mu_{e}$, for
which there is an extremal horizon\footnote{The case $z=4$ can be integrated
from scratch and it requires to impose $\xi=0$ in the self-interaction
(\ref{potencial1}). The metric function in this case reduces to
(\ref{gparalif1}), without the last term.}. For $\mu$ in the range $\mu
_{e}<\mu<0$, the topological Lifshitz black hole has an event and a Cauchy
horizon. When this upper bound is saturated, the Cauchy horizon shrinks to
cero, leading to a null singularity hidden by an event horizon. For $\mu>0$,
the spacetime has a single horizon, and its causal structure coincides with
that of Schwarzschild-AdS.

\bigskip

- For $z>4$ and $\xi=0$, the structure of the black holes in terms of the
different values of $\mu$ is exactly as the previous one. Nevertheless, for
$z>4$ and $\xi>0$, the analysis changes. There is a minimum value of $\mu$,
that can have either sign, for which the black hole is extremal, and above
which the are both an event and a Cauchy horizon.

\bigskip

Even though the range $1<z<2$ leads to a potential (\ref{potencial1}) that is
unbounded from below, there is a rich set of causal structures that are worth
to be analyzed even when one maintains the restriction $\xi>0$. In order to
explore the parameter space, it is useful to consider the plane $\left(
r_{h},\mu\right)  $ as well as the plane $\left(  \mu,\xi\right)  $, where
$r_{h}$ as a function of $\mu$ is obtained by solving $g\left(  r_{h}\right)  =0$, with
$g(r)$ given in (\ref{gparalif1}).

\begin{figure}
\centering
\includegraphics[scale=0.18]{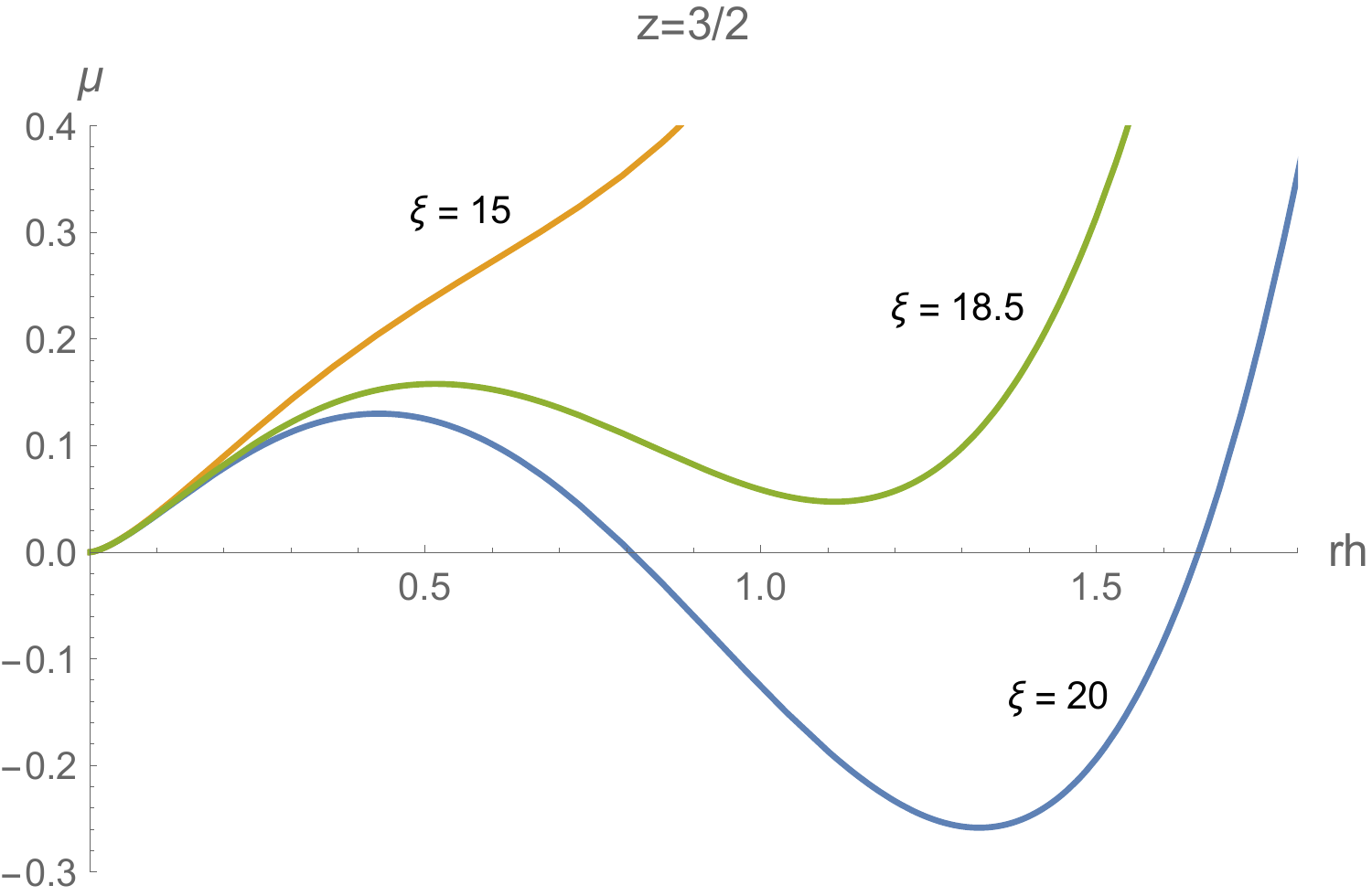}\qquad \includegraphics[scale=0.19]{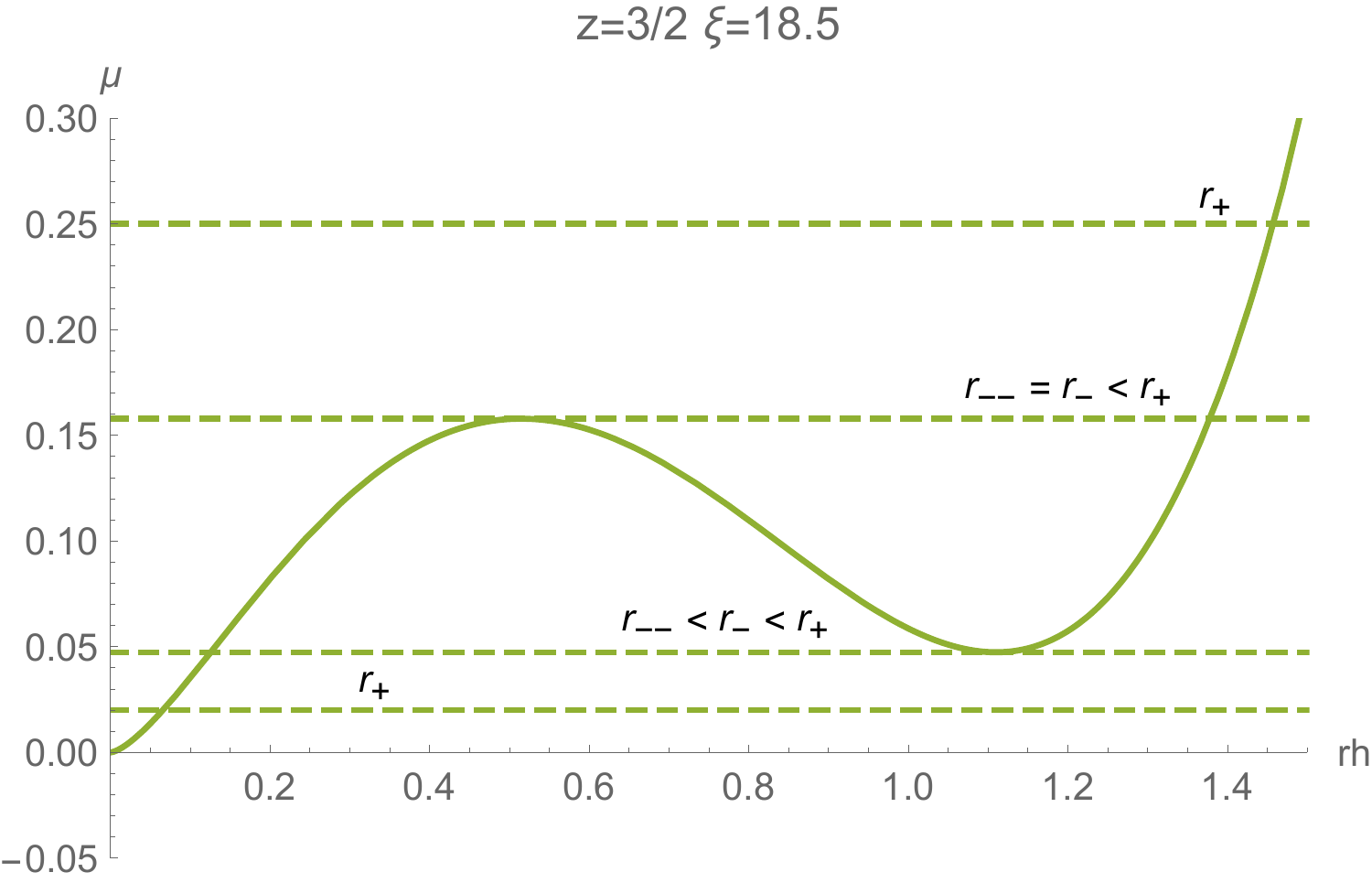}\qquad \includegraphics[scale=0.19]{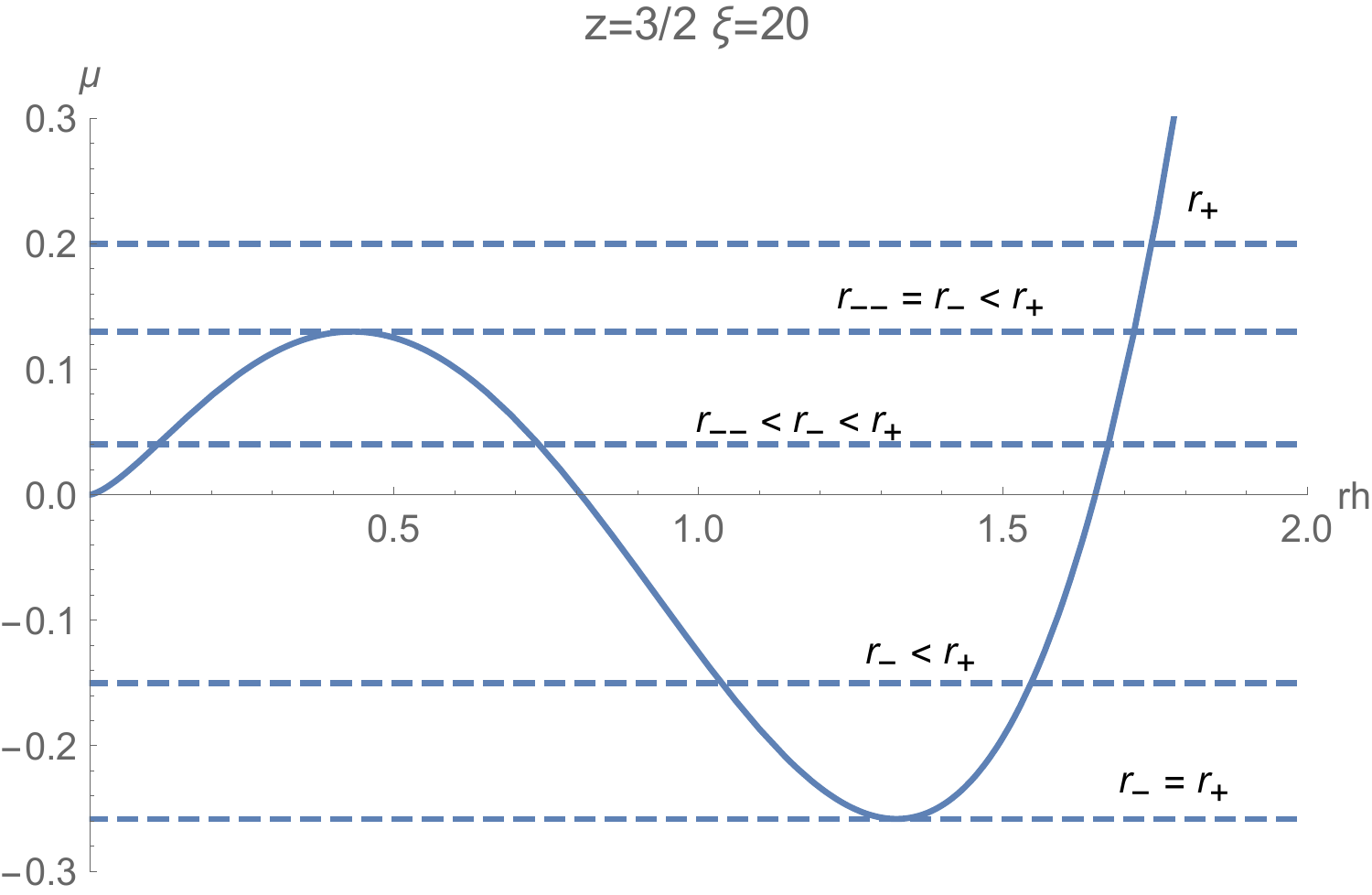}
 \caption{The panels show the structure of the curve $\mu=\mu(r_h)$ for different values of $\xi$ for $z=3/2$. For a given fixed value of $\mu$ the horizons are located at the intersections of the horizontal lines with the curves.}
    \label{ztremediosmuderh}
\end{figure}

In Figure \ref{ztremediosmuderh}, we have chosen $z=3/2$ for the three panels, that depict the
curves $\mu=\mu\left(  r_{h}\right)  $ for different values of the
self-coupling $\xi$. For a given value of the integration constant $\mu$, the
horizons will be located at the crossings of the horizontal line $\mu
=\mu_{\text{cte}}$ and the curves. In the first panel we see that there is a
critical value of $\xi=\xi_{crit}^{1}\sim17.2\sqrt{\frac{e}{l^3}}$, below
which there is a single event horizon for any value of $\mu>0$ (see e.g. the
curve with $\xi=15$). For $\xi_{crit}^{1}<\xi<\xi_{crit}^{2}\sim
18.78\sqrt{\frac{e}{l^{3}}}$ (see an example in the second panel), decreasing $\mu$ from infinity we have a black
hole with a single horizon, which for a given value of $\mu$ develops a
degenerate inner horizon, still surrounded by the external event horizon.
Then, there is a range of values for $\mu$ that lead to a black hole inside a
black hole structure. Such range for $\mu$ is bounded from below by a critical
value that leads to an inner horizon covered by a degenerate external event
horizon. Finally, for lower values of $\mu>0$, one recovers causal structure
of a unique event horizon. The second and third panels indicate the horizon
structures for $\xi=18.5$ and $\xi=20$, respectively.%

\begin{figure}
    \centering
    \includegraphics[scale=0.28]{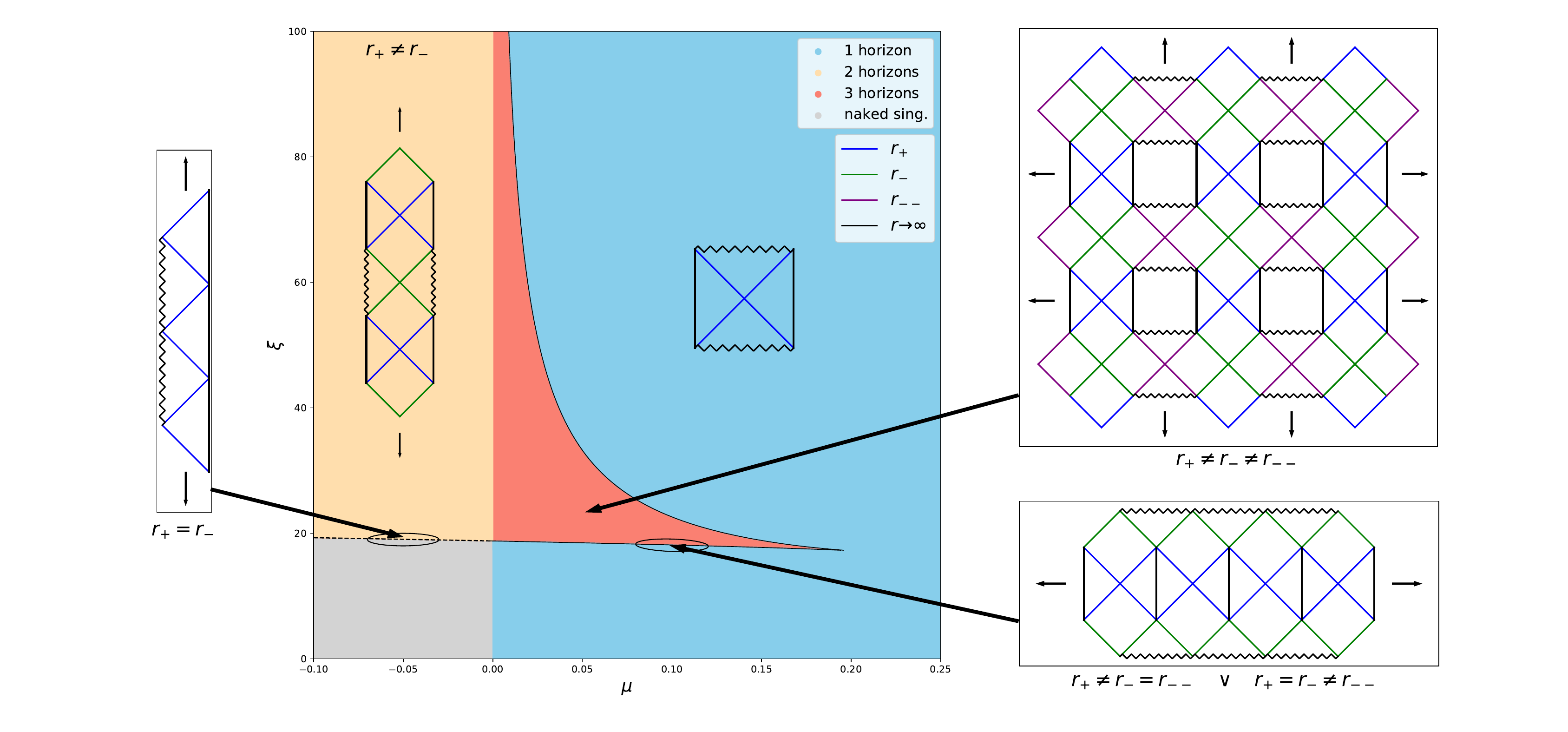}
    \caption{Causal structures that emerge for different values of the parameter $\xi$ and the integration constant $\mu$ for $z=3/2$.}
    \label{superfigura}
\end{figure}

Finally, Figure \ref{superfigura} summarizes the causal structures that can be obtained, in
the plane $\left(  \mu,\xi\right)  $ for $z=3/2$.
\\

\begin{figure}[h]
    \centering
    \includegraphics[scale=0.43]{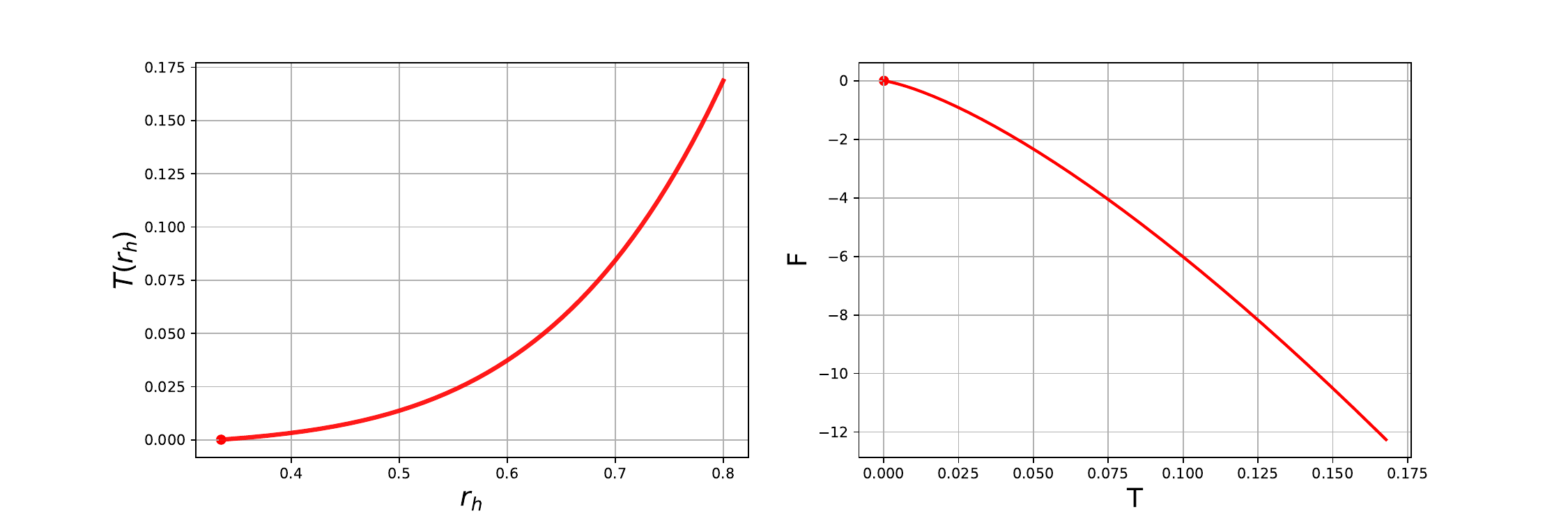}
    \caption{Temperature vs radius and Free energy vs temperature for black holes with $z=5,\,\xi=15\,\ell=1,\,e=1$.}
    \label{zmayor4}
\end{figure}

It is also worth to discuss the thermal properties of the black holes we have
considered in this section. As usual, the temperature of the event horizon can
be computed requiring the regularity of the Euclidean continuation. The
entropy can be computed using Wald's formula, and since all the fields are
regular on the horizon, and the couplings of matter with the Einstein-Hilbert
action are minimal, the entropy reduces to the Bekenstein-Hawking formula,
which in the normalization of the action (\ref{actionVgen1}) takes its usual
form%
\begin{equation}
S=\frac{A}{4G}=\frac{\pi r_{+}^{2}}{G}\ .
\end{equation}
Notice that the black holes that we are currently discussing are characterized
by a single integration constant, that defines the energy content of the
spacetime. One can obtain the mass of these black holes by the Abbott-Deser
method, adapted to asymptotically, topologically Lifshitz black holes, which
in this case leads to a finite result. In order to fix the absolute value of
the energy using the Abbott-Deser method one requires to identify a
background, nevertheless the first law will be fulfilled for any choice of
such reference geometry. As discusses above, for $z>2$ in order to have an
event horizon, the value of $\mu$ is always bounded from below by an extremal
case, which can be naturally chosen as a background metric (see e.g.
\cite{Tarrio:2011de}). One can show that for $z>2$, for a given choice of the temperature, there is a unique branch of black holes, which are always thermally favoured with respect to the extremal background since the free energy of the latter is set to zero, while the free energy of the former is always negative. Figure \ref{zmayor4} presents the corresponding plots.

\begin{figure}[h]
    \centering
    \includegraphics[scale=0.43]{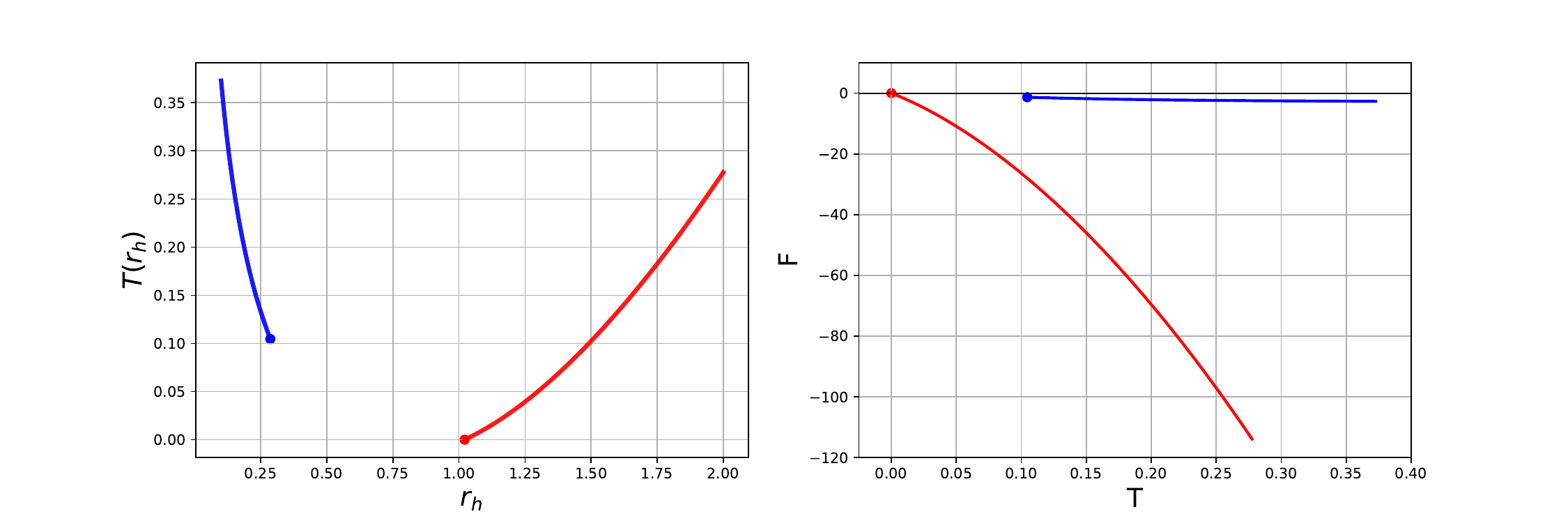}
    \caption{Temperature vs radius, and Free energy vs temperature for black holes with $z=3/2,\,\xi=18\,\ell=1,\,e=1$.}
    \label{thegappedbranches}
\end{figure}

\begin{figure}[h]
    \centering
    \includegraphics[scale=0.43]{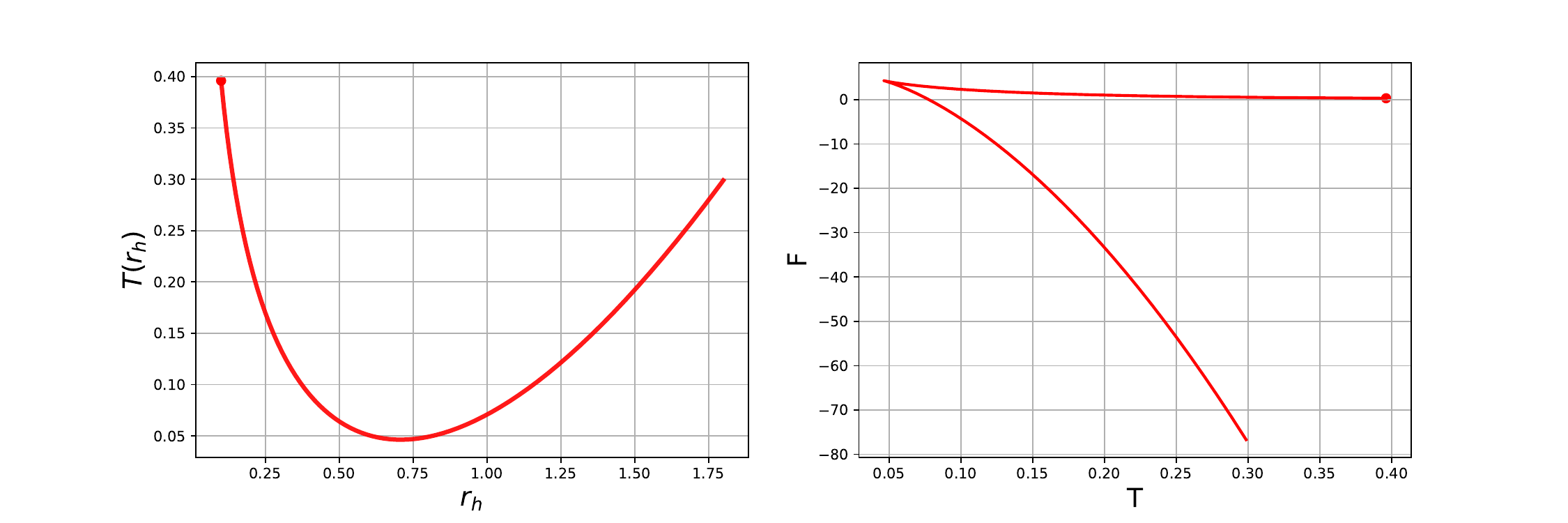}
    \caption{The figures show the temperature vs radius and Free energy vs temperature for black holes with $z=3/2$, $\xi=15$ and $l=1$. There is no extremal black hole in this family. Thus, the second panel only leads to sensible information if one is comparing the free energy between the large and the small black holes. In this case the configuration with vanishing free energy is actually a naked singularity.}
    \label{ztresmedionogap}
\end{figure}

The situation for $1<z<2$ is more subtle,
since there is a range of $\xi$ for which no extremal black hole exists, and
therefore there is no natural choice of background in this case. In Figures \ref{thegappedbranches} and \ref{ztresmedionogap} we have considered the thermal properties of event horizons $\xi=18\ell$ and $\xi=15\ell$, respectively, and fixing $z=3/2$. The former family contains an extremal black holes while the latter does not.

Let us discuss in some detail the case with the $z=\frac{3}{2}$ and $\xi=18\ell$, which is particularly interesting (see Figure \ref{thegappedbranches}). As the mass decreases the radius of the event horizon shrinks and eventually
the horizon becomes extremal at zero mass. If the mass falls below this minimum value, we still find a small black hole, which has a maximum size that
is gaped with respect to radius of the extremal black hole (see also panel two
of Figure \ref{ztremediosmuderh}). Large black holes, with radius above the extremal value always
dominate the canonical ensemble as depicted in Figure \ref{thegappedbranches}. The free energy of
the extremal case has been set to zero since we have used such geometry as background.

Some details of the case $z=3/2$ and $\xi=15$ are discussed in the caption of figure \ref{ztresmedionogap}, a family that does not contain an extremal black holes.

\bigskip

\subsection{Linear times exponential potentials}

The second family of potentials are given by a deformation of the scalar
potential that appears in $\mathcal{N}=4$ $SU\left(  2\right)  \times
SU\left(  2\right)  $ gauged supergravity (\ref{Vsugra}). In this section, in
order to present the exact solutions in a simple manner, we will see that it
is useful to normalize the action functional as follows%
\begin{equation}
I\left[  g_{\mu\nu},A,\phi\right]  =\int d^{4}x\sqrt{-g}\left(  \frac{R}%
{4}-\frac{1}{2}\partial_{\mu}\phi\partial^{\mu}\phi-\frac{e^{2\phi}}{2g^{2}%
}F_{\mu\nu}^{i}F^{i\mu\nu}-V_{2}\left(  \phi\right)  \right)  \label{defsugra}%
\end{equation}
with%
\begin{equation}
V_{2}\left(  \phi\right)  =\xi e^{-2\phi}+\zeta\phi e^{-2\phi}\ ,
\label{potdefsugra}%
\end{equation}
and%
\begin{equation}
\zeta=\frac{\left(  g^{2}\eta^{2}-1\right)  }{\eta^{4}g^{2}}\ ,
\end{equation}
which has to be interpreted as an equation for $\eta$ in terms of the coupling
$g$ and $\zeta$. The potential (\ref{potdefsugra}) is bounded from below when
$\zeta>0$ for either sign of $\xi$, while for $\zeta<0$ the potential is
unbounded from below. Notice that we have fixed the dilatonic coupling, which
is required if we want to consider this theory as a deformation of a
truncation of the mentioned supergravity theory, that is obtained by modifying
the potential only. Indeed, when $\zeta=0$, the theory (\ref{defsugra}) is
recovered as a consistent truncation of $\mathcal{N}=4$ $SU\left(  2\right)  \times
SU\left(  2\right)  $ gauged supergravity by setting $\xi=-\frac{g^{2}}{8}$
and $\eta=1/g$. Again,
the gauge field is given by the meron configuration (\ref{meronVs}) while in
this case the dilaton reduces to%
\begin{equation}
\phi\left(  r\right)  =\ln\left(  \frac{r}{\eta}\right)  \ .
\end{equation}
For the present family of solutions, the metric of the spacetime is given by%
\begin{equation}
ds^{2}=-r^{2}f\left(  r\right)  dt^{2}+\frac{dr^{2}}{f\left(  r\right)
}+r^{2}d\Omega^{2}\ ,
\end{equation}
with%
\begin{equation}
f\left(  r\right)  =-\zeta\ln\left(  \frac{r}{\eta}\right)  \eta^{2}+\left(
\zeta-\xi\right)  \eta^{2}+\frac{1}{4\eta^{2}g^{2}}-\frac{\mu}{r^{2}}\ .
\label{fdedlV3}%
\end{equation}
Here $\mu$ is an integration constant. In spite of the fact that the metric
has a logarithmically growing term in the areal radial coordinate, the
curvature of the spacetime vanishes when $r\rightarrow\infty$, therefore the
spacetime is asymptotically locally flat.

Using the Abbott-Deser method, and choosing as a background the spacetime with
$\mu=0$, leads to the following expression for the mass (cf. equation
(\ref{defsugra})):%
\begin{equation}
M=2\pi\mu\ .
\end{equation}

Now we proceed to analyze the causal structures contained in this family of
black holes. The equation for the zeros of the function $f\left(  r\right)  $
in (\ref{fdedlV3}) leads to the a Lambert W function\footnote{The Lambert W
function is defined by $W\left(  z\right)  e^{W\left(  z\right)  }=z$. This
relation leads to a multivalued $W=W\left(  z\right)  $. Restricting our
attention to $W\left(  z\right)  :\mathbb{R}\rightarrow \mathbb{R}$, there are two branches
$W_{0}\left(  z\right)  :\left(  -e^{-1},\infty\right)  \rightarrow\left(
-1,+\infty\right)  $ and $W_{1}:\left(  -e^{-1},0\right)  \rightarrow\left(
-1,-\infty\right)  $.}. For $\zeta>0$ the largest zero of (\ref{fdedlV3})
represents a cosmological horizon, while for $\zeta<0$ the largest zero of
(\ref{fdedlV3}) gives rise to an event horizon. When $\zeta$ vanishes there
could be only one horizon located at%
\begin{equation}
\left.  r_{h}\right\vert _{\zeta=0}=\left(  \frac{4\mu g^{2}}{g^{2}-4\xi
}\right)  ^{\frac{1}{2}}\ ,
\end{equation}
which for $g^{2}>4\xi$ and $\mu>0$ is an event horizon, while for $g^{2}<4\xi$
and $\mu<0$ represents a cosmological horizon.

Now, let us further analyze the Lambert W function obtained for the location
of the would be horizons, which will be useful in the study of the thermal
properties of these black holes. Indeed, the horizons will be located at
$r=r_{h}$ such that%
\begin{equation}
f\left(  r\right)  =a+br_{h}^{2}\ln r+cr_{h}^{2}=0\ ,
\end{equation}
which setting $x=r_{h}^{-2}$ leads to
\begin{equation}
ax-\frac{b}{2}\ln x+c=0\ , \label{casilambert}%
\end{equation}
where%
\begin{equation}
a=-\mu,\text{ }b=-\zeta\eta^{2},\ c=\frac{1}{4\eta^{2}g^{2}}+\left(  \zeta
-\xi\right)  \eta^{2}+\zeta\eta^{2}\ln\eta\ .
\end{equation}
The existence, and number of solutions of (\ref{casilambert}) depends on the
value of the discriminant\footnote{See reference \cite{BravoGaete:2019rci} for an example of a different setup where the location of the event horizon of a black hole is given in terms of a Lambert W function. The formulae in the appendix of such reference are relevant to our analysis.}%
\begin{equation}
\Delta=-2\frac{a}{b}e^{2\frac{c}{b}}\ .
\end{equation}
We have two zeros when $-\frac{1}{e}<\Delta<0$, which restricts the values of
$\mu$ depending also on the sign of $\zeta$ (see analysis below). Within such
ranges the two roots of (\ref{casilambert}) are given by
\begin{equation}
r_{h_{2}}=e^{\frac{1}{2}W_{0}\left(  \Delta\right)  -\frac{c}{b}},\qquad
r_{h_{1}}=e^{\frac{1}{2}W_{-1}\left(  \Delta\right)  -\frac{c}{b}}\ ,
\end{equation}
and since $W_{0}\left(  x\right)  >W_{-1}\left(  x\right)  ,\ x\in\left(
-\frac{1}{e},0\right)  $, the root $r_{h_{2}}$ is always larger than
$r_{h_{1}}$. For $\zeta>0$ in order to have two roots,$\mu$ must lie within
the range $0<\mu<\mu_{\max}=\frac{\eta^{2}\zeta}{2}e^{1+\frac{2c}{b}}$, and as
mentioned above $r_{h_{2}}=r_{++}$ and $r_{h_{1}}=r_{+}$ correspond to a
cosmological and an event horizon, respectively. On the other hand, for
$\zeta<0$ and $\mu_{\min}=-\frac{\eta^{2}|\zeta|}{2}e^{1+\frac{2c}{b}}<\mu<0$,
there are two horizons as well, located at $r_{h_{2}}=r_{+}$ and $r_{h_{1}%
}=r_{-}$, which in this case correspond to an event and a Cauchy horizon, respectively.

For $\Delta\geq0$ and for $\Delta=-\frac{1}{e}$, there is a unique horizon,
which can be an event or a cosmological horizon, depending on the values of
the remaining constants. Finally for $\Delta<-e^{-1}$, there are no horizon.

\bigskip

Let us now provide the expression for the temperature and the entropy for a
horizon located at $r=r_{\star}$. The former reads%
\begin{align}
T_{\star}  &  =\frac{1}{4\pi}\left.  \sqrt{\frac{d}{dr}\left(  r^{2}\left(
\frac{a}{r^{2}}+b\ln r+c\right)  \right)  \frac{d}{dr}\left(  \left(  \frac
{a}{r^{2}}+b\ln r+c\right)  \right)  }\right\vert _{r=r_{\star}}\\
&  =\frac{\sqrt{b}}{4\pi}\sqrt{b+bW_{\star}\left(  \Delta\right)
-2ae^{-W_{\star}\left(  \Delta\right)  +2\frac{c}{b}}-2aW_{\star}\left(
\Delta\right)  e^{-W_{\star}\left(  \Delta\right)  +2\frac{c}{b}}}\ ,
\label{temperaturelambert}%
\end{align}
while the entropy, is given by the Bekenstein-Hawking formula, which in the
normalization of (\ref{defsugra}) leads to
\begin{equation}
S_{\star}=4\pi^{2}e^{W_{\star}\left(  \Delta\right)  -\frac{2c}{b}}\ .
\end{equation}
Here $W_{\star}\left(  \Delta\right)  $ has to be understood as $W_{-1}\left(
\Delta\right)  $ or $W_{0}\left(  \Delta\right)  $ depending on whether we are
dealing with a single event horizon, a cosmological and an event
horizon, or an event and a Cauchy horizon.

\bigskip

As a final consistency check, let us verify that the first law is fulfilled,
which will turn out to require using some identities satisfied by the Lambert
W function. Our black holes are characterized by a single integration
constant, $\mu$, and the difference in entropy of two equilibrium
configurations corresponding to the values $\mu$ and $\mu+\delta\mu$, is given
by%
\begin{align}
\delta S_{\star}  &  =4\pi^{2}e^{W_{\star}\left(  \Delta\right)  -\frac{2c}%
{b}}\delta\left(  W_{\star}\left(  \Delta\right)  \right) \\
&  =4\pi^{2}e^{W_{\star}\left(  \Delta\right)  -\frac{2c}{b}}W_{\star}%
^{\prime}\left(  \Delta\right)  \frac{2\delta\mu}{b}e^{2\frac{c}{b}}\\
&  =-4\pi^{2}\frac{e^{W_{\star}\left(  \Delta\right)  -\frac{2c}{b}}}%
{\Delta+e^{W_{\star}\left(  \Delta\right)  }}\frac{\Delta}{a}\delta\mu\ ,
\end{align}
where in going from the before to last to the last line we have used the
following identity of the derivative of the Lambert W function:%
\begin{equation}
W_{\star}^{\prime}\left(  \Delta\right)  =\left(  \Delta+e^{W_{\star}\left(
\Delta\right)  }\right)  ^{-1}\ .
\end{equation}
Multiplying $\delta S_{\star}$ times the temperature (\ref{temperaturelambert}%
) leads to%
\begin{align}
T_{GH}\delta S_{\star}  &  =-4\pi^{2}\delta\mu be^{-2\frac{c}{b}}\frac{\Delta
}{a}\frac{\sqrt{b}}{4\pi}\sqrt{\frac{1+W_{\star}\left(  \Delta\right)
}{b+\frac{W_{\star}\left(  \Delta\right)  }{\Delta}b\left(  -2\frac{a}%
{b}e^{\frac{2c}{b}}\right)  }}\\
&  =-\pi\delta\mu b\frac{1}{a}\left(  -2\frac{a}{b}e^{2\frac{c}{b}}\right)
e^{-2\frac{c}{b}}\\
&  =2\pi\delta\mu
\end{align}
which shows that the first law is fulfilled for every potential type of
horizon%
\begin{equation}
\delta M=T_{-}\delta S_{-}\text{ and }\delta M=T_{+}\delta S_{+}\text{ and
}\delta M=T_{++}\delta S_{++}\ .
\end{equation}

\section{Conclusions}

In this work, we have started by revisiting the problem of BPS solutions in
the $\mathcal{N}=4$ $SU\left(  2\right)  \times SU\left(  2\right)  $ gauged
supergravity. In the Abelian sector of gauge fields of the theory, we have found a new completely regular, soliton spacetime, which preserves one-quarter of the supersymmetry. The soliton is charged, and asymptotically admits an extra conformal Killing vector. This spacetime can also be obtained from the double analytic continuation of a plannar solution found in \cite{Klemm}. The BPS conditions lead to a naked singularity in \cite{Klemm}, nevertheless due to the double Wick rotation, the conditions for unbroken supersymmetry in our case lead to a regular spacetime. Also in the Abelian sector of the theory, but now assuming spherical symmetry on the metric, we have
also found a new BPS configuration which preserves one-quarter of the
supersymmetries, and that describes a naked singularity. Notice that singular
spacetimes are known to appear as BPS solutions in gauged supergravities
\cite{Romans}.

The supergravity theory considered in this work has a potential for
the dilaton without a local extremum, which leads to asymptotically locally flat spacetimes, instead of locally AdS as
it is the case in other gauged supergravities. Then, we have moved to the
construction of new, non-Abelian solutions, by considering the meron
ansatz \cite{deAlfaro:1976qet}. Since the supergravity theory contains two $su\left(  2\right)
$ gauge fields, we have constructed electric-meronic solutions as well as
double-meron solutions with a spacetime that is spherically symmetric. The
latter leads to one-quarter BPS configurations where the spacetime is again
singular. It is interesting to remark that the non-BPS black holes in the double-meron
sector, have a temperature that is independent of the mass of the spacetime,
which can be seen as a signal of criticality. It would be interesting to allow
non-trivial profiles for the axion, which has been set to zero along our work,
and to explore the construction of more general exact solutions in this supergravity model.

A thorough exploration of self-interactions that allow for the explicit
construction of black holes has proven to be a worth task in different contexts,
since for example in the series of works \cite{Anabalon:2012ta}-\cite{Acena:2013jya} such analysis turned out
to lead to a one-parameter deformation of the four, single scalar truncations of the maximal supergravity in four dimensions \cite{deWitNicolai}, as well as to the potentials of the two cases that admit an $\omega$-deformation \cite{omega} in the single dilaton consistent sectors identified in \cite{Tarrio:2013qga}.
With this in mind, in Section V we have moved beyond supergravity, but
keeping the metric, one gauge field and the dilaton as field content of the
theory. Within the meron ansatz for the gauge field, we have found that there
are at least two families of self-interactions for the scalar field which
allow to find exact analytic black holes with interesting properties. The
first family of self-interactions leads to topologically Lifshitz black holes \cite{Mann:2009yx} with a variety of causal structures, containing for
example a black hole inside a black hole.
The second family of potentials can be seen as a two-parameter deformation of
the dilaton potential of the Freedman-Schwarz model. The spacetime metric
contains a logarithmically growing term, in spite of which the asymptotic region is locally flat. In this case there can be a single event horizon,
a single cosmological horizon, an event horizon surrounded by a cosmological
one, and finally an event horizon hiding a Cauchy horizon; the latter
configuration can achieve extremality.

Our work shows that the meron ansatz provides an interesting arena for the
construction of new solutions in gauged supergravity (see also \cite{Bueno:2014mea}). It is know that $\mathcal{N}=4$ supergravity admits as
well a different gauging, which
leads to gauge fields valued on $SO\left(  4\right)$ \cite{Gates:1982ct}, and to a potential for the self-interaction that can have a
local extremum. It would be interesting to explore the compatibility of the meron ansatz in the $\mathcal{N}=4$ $SO\left(  4\right)  $ gauged theory. Work along these lines is in progress. Another interesting future direction is to consider a ten-dimensional embedding. Indeed, one may hope that there could be an effective $SL(2,R)$ which will acts on the torus contained in\footnote{We thank the anonymous referee for this illuminating comment} $S^3\times S^3$.

\section{Acknowledgments} 
We thank Andrés Anabalón, Cristóbal Corral, José Figueroa and Carlos Nuñez, for valuable comments and discussions. This work is partially funded by Beca ANID de Magíster 22201618 and FONDECYT grants 1181047 and 1200022. J.O. also thanks the support of Proyecto de Cooperaci\'on Internacional 2019/13231-7 FAPESP/ANID. The Centro de Estudios Cient\'{\i}ficos (CECs) is funded by the Chilean
Government through the Centers of Excellence Base Financing Program of ANID.

\appendix
\normalsize
\section{Explicit expressions for the Killing spinors of Sections
II and III and IV}

\textbf{Killing spinors for the Soliton}: The Killing spinors for the soliton presented in Section II are given by
\footnotesize
\begin{equation*}
\bar{\epsilon}_{1}=\left( \cosh l\right) ^{-1/4}\left( 
\begin{array}{c}
-\sqrt{e_{A}^{2}+e_{B}^{2}}\sqrt{\cosh l-1} \\ 
\sqrt{\cosh l+1} \\ 
0 \\ 
\frac{ie_{A}}{e_{B}}\sqrt{\cosh l+1} \\ 
i\frac{\sqrt{e_{A}^{2}+e_{B}^{2}}}{e_{B}}\sqrt{\cosh l-1} \\ 
-i\sqrt{\cosh l+1} \\ 
0 \\ 
\frac{e_{A}}{e_{B}}\sqrt{\cosh l+1} \\ 
0 \\ 
0_{8\times 1}%
\end{array}%
\right) \ ,\ \bar{\epsilon}_{2}=\left( \cosh l\right) ^{-1/4}\left( 
\begin{array}{c}
i\sqrt{e_{A}^{2}+e_{B}^{2}}\sqrt{\cosh l+1} \\ 
i\sqrt{\cosh l-1} \\ 
0 \\ 
-\frac{e_{A}}{e_{B}}\sqrt{\cosh l-1} \\ 
-\frac{\sqrt{e_{A}^{2}+e_{B}^{2}}}{e_{B}}\sqrt{\cosh l+1} \\ 
-\sqrt{\cosh l-1} \\ 
0 \\ 
-i\frac{e_{A}}{e_{B}}\sqrt{\cosh l-1} \\ 
0 \\ 
0_{8\times 1}%
\end{array}%
\right) 
\end{equation*}%
\begin{equation*}
\bar{\epsilon}_{3}=\left( \cosh l\right) ^{-1/4}\left( 
\begin{array}{c}
i\frac{e_{A}}{e_{B}}\frac{\sqrt{\cosh l+1}}{{}} \\ 
0 \\ 
\sqrt{\cosh l+1} \\ 
\frac{\sqrt{e_{A}^{2}+e_{B}^{2}}}{e_{B}}\sqrt{\cosh l-1} \\ 
-\frac{e_{A}}{e_{B}}\sqrt{\cosh l+1} \\ 
0 \\ 
i\sqrt{\cosh l+1} \\ 
i\frac{\sqrt{e_{A}^{2}+e_{B}^{2}}}{e_{B}}\sqrt{\cosh l-1} \\ 
0 \\ 
0_{8\times 1}%
\end{array}%
\right) \ ,\ \bar{\epsilon}_{4}=\left( \cosh l\right) ^{-1/4}\left( 
\begin{array}{c}
-\frac{e_{A}}{e_{B}}\sqrt{\cosh l-1} \\ 
0 \\ 
i\sqrt{\cosh l-1} \\ 
-i\frac{\sqrt{e_{A}^{2}+e_{B}^{2}}}{e_{B}}\sqrt{\cosh l+1} \\ 
i\frac{e_{A}}{e_{B}}\sqrt{\cosh l-1} \\ 
0 \\ 
\sqrt{\cosh l-1} \\ 
-\frac{\sqrt{e_{A}^{2}+e_{B}^{2}}}{e_{B}}\sqrt{\cosh l+1} \\ 
0 \\ 
0_{8\times 1}%
\end{array}%
\right) 
\end{equation*}
\normalsize
which shows that the soliton spacetime preserves 1/4 of the supersymmetries of the theory.

\bigskip

\textbf{Killing spinors for the spherically symmetric, supersymmetric solution in the Abelian sector:} For the Abelian BPS solution discussed in section III, the four Killing spinors that preserve the
supersymmetry transformations have half of the components vanishing. It is not possible to factorize the dependence on the radial coordinate $\rho$. Nevertheless, the explicit form of the Killing spinors can be given in a compact manner as follows:
\begin{equation*}
\bar{\epsilon}_{1}=\Psi _{1}\left( \rho \right) \left( 
\begin{array}{c}
-e_{A}c_{\varphi }s_{\theta } \\ 
-ie_{A}s_{\varphi }s_{\theta } \\ 
\frac{2\Lambda }{\sin \theta }s_{\varphi }c_{\theta }^{2}s_{\theta
}+ie_{B}c_{\varphi }s_{\theta } \\ 
i\Lambda c_{\varphi }c_{\theta }-e_{B}s_{\varphi }s_{\theta } \\ 
\Lambda c_{\varphi }c_{\theta }+\frac{2ie_{B}}{\sin \theta }s_{\varphi
}s_{\theta }^{2}c_{\theta } \\ 
-\frac{2i\Lambda }{\sin \theta }s_{\varphi }c_{\theta }^{2}s_{\theta
}+e_{B}c_{\varphi }s_{\theta } \\ 
-\frac{2e_{A}}{\sin \theta }s_{\varphi }c_{\theta }s_{\theta }^{2} \\ 
ie_{A}c_{\varphi }s_{\theta } \\ 
0_{8\times 1}%
\end{array}%
\right) ^{T}+\Psi _{2}\left( \rho \right) \left( 
\begin{array}{c}
-e_{A}c_{\varphi }c_{\theta } \\ 
ie_{A}s_{\varphi }c_{\theta } \\ 
-\frac{2\Lambda }{\sin \theta }s_{\varphi }s_{\theta }^{2}c_{\theta
}+ie_{B}c_{\varphi }c_{\theta } \\ 
i\Lambda c_{\varphi }s_{\theta }+e_{B}s_{\varphi }c_{\theta } \\ 
-\Lambda c_{\varphi }s_{\theta }+\frac{i2e_{B}}{\sin \theta }s_{\varphi
}s_{\theta }c_{\theta }^{2} \\ 
-\frac{i2\Lambda }{\sin \theta }s_{\varphi }s_{\theta }^{2}c_{\theta
}-e_{B}c_{\varphi }c_{\theta } \\ 
-\frac{2e_{A}}{\sin \theta }s_{\varphi }c_{\theta }^{2}s_{\theta } \\ 
-ie_{A}c_{\varphi }c_{\theta } \\ 
0_{8\times 1}%
\end{array}%
\right) ^{T}
\end{equation*}%
\begin{equation*}
\bar{\epsilon}_{2}=\Psi _{1}\left( \rho \right) \left( 
\begin{array}{c}
-e_{A}c_{\theta }c_{\varphi } \\ 
ie_{A}s_{\varphi }c_{\theta } \\ 
\frac{2\Lambda }{\sin \theta }s_{\varphi }s_{\theta }^{2}c_{\theta
}+ie_{B}c_{\varphi }c_{\theta } \\ 
e_{B}s_{\varphi }c_{\theta }-i\Lambda c_{\varphi }s_{\theta } \\ 
-\Lambda c_{\varphi }s_{\theta }-\frac{2ie_{B}}{\sin \theta }s_{\varphi
}c_{\theta }^{2}s_{\theta } \\ 
-\frac{i2\Lambda }{\sin \theta }s_{\varphi }s_{\theta }^{2}c_{\theta
}+e_{B}c_{\varphi }c_{\theta } \\ 
\frac{2e_{A}}{\sin \theta }s_{\varphi }s_{\theta }c_{\theta }^{2} \\ 
ie_{A}c_{\varphi }c_{\theta } \\ 
0_{8\times 1}%
\end{array}%
\right) ^{T}+\Psi _{2}\left( \rho \right) \left( 
\begin{array}{c}
e_{A}c_{\varphi }s_{\theta } \\ 
ie_{A}s_{\theta }s_{\varphi } \\ 
\frac{2\Lambda }{\sin \theta }s_{\varphi }c_{\theta }^{2}s_{\theta
}-ie_{B}s_{\theta }c_{\varphi } \\ 
ic_{\varphi }c_{\theta }\Lambda +e_{B}s_{\theta }s_{\varphi } \\ 
-c_{\varphi }c_{\theta }\Lambda +\frac{2ie_{B}}{\sin \theta }s_{\varphi
}s_{\theta }^{2}c_{\theta } \\ 
\frac{2i\Lambda }{\sin \theta }s_{\varphi }c_{\theta }^{2}s_{\theta
}+e_{B}c_{\varphi }s_{\theta } \\ 
-\frac{2e_{A}}{\sin \theta }s_{\varphi }s_{\theta }^{2}c_{\theta } \\ 
ie_{A}c_{\varphi }s_{\theta } \\ 
0_{8\times 1}%
\end{array}%
\right) ^{T}
\end{equation*}%
\begin{equation*}
\bar{\epsilon}_{3}=\Psi _{1}\left( r\right) \left( 
\begin{array}{c}
e_{A}s_{\varphi }s_{\theta } \\ 
-e_{A}ic_{\varphi }s_{\theta } \\ 
\frac{2\Lambda }{\sin \theta }c_{\varphi }c_{\theta }^{2}s_{\theta
}-ie_{B}s_{\varphi }s_{\theta } \\ 
-e_{B}c_{\varphi }s_{\theta }-i\Lambda s_{\varphi }c_{\theta } \\ 
-\Lambda s_{\varphi }c_{\theta }+\frac{i2e_{B}}{\sin \theta }c_{\varphi
}s_{\theta }^{2}c_{\theta } \\ 
-\frac{i2\Lambda }{\sin \theta }c_{\varphi }c_{\theta }^{2}s_{\theta
}-e_{B}s_{\varphi }s_{\theta } \\ 
-\frac{2e_{A}}{\sin \theta }c_{\varphi }c_{\theta }s_{\theta }^{2} \\ 
-ie_{A}s_{\varphi }s_{\theta } \\ 
0_{8\times 1}%
\end{array}%
\right) ^{T}+\Psi _{2}\left( r\right) \left( 
\begin{array}{c}
e_{A}s_{\varphi }c_{\theta } \\ 
ie_{A}c_{\varphi }c_{\theta } \\ 
-\frac{2\Lambda }{\sin \theta }c_{\varphi }s_{\theta }^{2}c_{\theta
}-ie_{B}s_{\varphi }c_{\theta } \\ 
e_{B}c_{\varphi }c_{\theta }-i\Lambda s_{\varphi }s_{\theta } \\ 
\Lambda s_{\varphi }s_{\theta }+\frac{i2e_{B}}{\sin \theta }c_{\varphi
}c_{\theta }^{2}s_{\theta } \\ 
-\frac{i2\Lambda }{\sin \theta }c_{\varphi }s_{\theta }^{2}c_{\theta
}+e_{B}s_{\varphi }c_{\theta } \\ 
-\frac{2e_{A}}{\sin \theta }c_{\varphi }c_{\theta }^{2}s_{\theta } \\ 
ie_{A}s_{\varphi }c_{\theta } \\ 
0_{8\times 1}%
\end{array}%
\right) ^{T}
\end{equation*}%
\begin{equation*}
\bar{\epsilon}_{4}=\Psi _{1}\left( \rho \right) \left( 
\begin{array}{c}
-e_{A}s_{\varphi }c_{\theta } \\ 
-ie_{A}c_{\varphi }c_{\theta } \\ 
-\frac{2\Lambda }{\sin \theta }c_{\varphi }s_{\theta }^{2}c_{\theta
}+ie_{B}c_{\theta }s_{\varphi } \\ 
-i\Lambda s_{\theta }s_{\varphi }-c_{\varphi }e_{B}c_{\theta } \\ 
-\Lambda s_{\varphi }s_{\theta }+\frac{i2e_{B}}{\sin \theta }c_{\varphi
}c_{\theta }^{2}s_{\theta } \\ 
\frac{i2\Lambda }{\sin \theta }c_{\varphi }s_{\theta }^{2}c_{\theta
}+e_{B}s_{\varphi }c_{\theta } \\ 
-\frac{2e_{A}}{\sin \theta }c_{\varphi }c_{\theta }^{2}s_{\theta } \\ 
ie_{A}s_{\varphi }c_{\theta } \\ 
0_{8\times 1}%
\end{array}%
\right) ^{T}+\Psi _{2}\left( \rho \right) \left( 
\begin{array}{c}
e_{A}s_{\varphi }s_{\theta } \\ 
-ie_{A}c_{\varphi }s_{\theta } \\ 
-\frac{2\Lambda }{\sin \theta }c_{\varphi }c_{\theta }^{2}s_{\theta
}-ie_{B}s_{\varphi }s_{\theta } \\ 
i\Lambda s_{\varphi }c_{\theta }-e_{B}c_{\varphi }s_{\theta } \\ 
-\Lambda s_{\varphi }c_{\theta }-\frac{i2e_{B}}{\sin \theta }c_{\varphi
}s_{\theta }^{2}c_{\theta } \\ 
-\frac{i2\Lambda }{\sin \theta }c_{\varphi }c_{\theta }^{2}s_{\theta
}+e_{B}s_{\varphi }s_{\theta } \\ 
\frac{2e_{A}}{\sin \theta }c_{\varphi }s_{\theta }^{2}c_{\theta } \\ 
ie_{A}s_{\theta }s_{\varphi } \\ 
0_{8\times 1}%
\end{array}%
\right) ^{T}
\end{equation*}
where%
\begin{eqnarray*}
\Psi _{1}\left( \rho \right)  &=&\frac{\sqrt{\left( \rho
e_{A}+i2Q_{A}\right) \Lambda ^{2}H_{A}-ie_{A}e_{B}\rho }}{\rho ^{1/4}} \text{ ,}\\
\Psi _{2}\left( \rho \right)  &=&\frac{e_{A}e_{B}\sqrt{f_{BPS}\left( \rho \right)
\rho }}{\rho ^{1/4}\sqrt{\left( \rho e_{A}+i2Q_{A}\right) \Lambda
^{2}H_{A}-ie_{A}e_{B}\rho }}\text{ ,}
\end{eqnarray*}
and $f_{BPS}(\rho)$ is defined by \eqref{fBPS}.

\bigskip

\textbf{Killing spinors for the double-meron:} In section IV we have shown that the double-meron solution with $\mu=0$
admits a set of four Killing spinors that satisfy the equations
(\ref{unmedio-transf}) and (\ref{RS-transf}). Now we will provide the explicit form of these spinors, which take the form
\begin{align}
    \bar{\epsilon}_{i}=\sqrt{r}\left(  A\otimes B_{i}+\eta\otimes C_{i}\right)
^{T}%
\end{align}
where $T$ means transpose and the
vectors $A$ and $\eta$ are common for the four spinors and are given by
\[
A=\left(
\begin{array}
[c]{c}%
\sin\theta\cos\varphi\\
\sin\theta\sin\phi\\
\cos\theta\\
0
\end{array}
\right)  \ \text{,}\ \eta=\left(
\begin{array}
[c]{c}%
0\\
0\\
0\\
1
\end{array}
\right)  \ .
\]
Notice that the radial dependence is factorized on the killing spinor and the
depends on the angles are in the vector and spinors $A,\,B_{i}\text{ and }%
C_{i}$ which are given by $\left( \tilde{\Lambda}\rightarrow \tilde{\Lambda}%
2e_{A}e_{B}\right) $\newline
\begin{equation*}
B_{1}=\left( 
\begin{array}{c}
c_{\varphi }s_{\theta } \\ 
is_{\varphi }s_{\theta } \\ 
\frac{2e_{B}\tilde{\Lambda}}{\Lambda }s_{\varphi }s_{\theta }+i\frac{e_{B}}{%
\Lambda }s_{\varphi }c_{\theta }-i\frac{e_{B}}{e_{A}}c_{\varphi }s_{\theta }
\\ 
-i\frac{2e_{B}\tilde{\Lambda}}{\Lambda }\left( c_{\varphi }s_{\theta
}\right) -\frac{e_{B}}{\Lambda }c_{\varphi }c_{\theta }+\frac{e_{B}}{e_{A}}%
s_{\varphi }s_{\theta }%
\end{array}%
\right) \ ,\ C_{1}=\left( 
\begin{array}{c}
i\frac{2e_{B}^{2}\tilde{\Lambda}}{\Lambda ^{2}}s_{\varphi }c_{\theta }+\frac{%
e_{B}^{2}}{\Lambda ^{2}}s_{\varphi }s_{\theta }+\frac{\Lambda }{e_{A}}%
c_{\varphi }c_{\theta } \\ 
-\frac{2e_{B}^{2}\tilde{\Lambda}}{\Lambda ^{2}}c_{\varphi }c_{\theta }-i%
\frac{e_{B}^{2}}{\Lambda ^{2}}c_{\varphi }s_{\theta }-i\frac{\Lambda }{e_{A}}%
s_{\varphi }c_{\theta } \\ 
\frac{4e_{A}e_{B}}{\Lambda ^{2}\sin ^{2}\theta }s_{\theta }^{2}c_{\theta
}^{2}\left( is_{\varphi }s_{\theta }-2\tilde{\Lambda}s_{\varphi }c_{\theta
}\right)  \\ 
\frac{4e_{A}e_{B}}{\Lambda ^{2}\sin ^{2}\theta }s_{\theta }^{2}c_{\theta
}^{2}\left( c_{\varphi }s_{\theta }-2i\tilde{\Lambda}c_{\varphi }c_{\theta
}\right) 
\end{array}%
\right) 
\end{equation*}%
\begin{equation*}
B_{2}=\left( 
\begin{array}{c}
c_{\varphi }c_{\theta } \\ 
-is_{\varphi }c_{\theta } \\ 
-\frac{2e_{B}\tilde{\Lambda}}{\Lambda }s_{\varphi }c_{\theta }+i\frac{e_{B}}{%
\Lambda }s_{\varphi }s_{\theta }-i\frac{e_{B}}{e_{A}}c_{\varphi }c_{\theta }
\\ 
-i\frac{2e_{B}\tilde{\Lambda}}{\Lambda }c_{\varphi }c_{\theta }+\frac{e_{B}}{%
\Lambda }c_{\varphi }s_{\theta }-\frac{e_{B}}{e_{A}}s_{\varphi }c_{\theta }%
\end{array}%
\right) \ ,\ C_{2}=\left( 
\begin{array}{c}
i\frac{2e_{B}^{2}\tilde{\Lambda}}{\Lambda ^{2}}s_{\varphi }s_{\theta }-\frac{%
e_{B}^{2}}{\Lambda ^{2}}s_{\varphi }c_{\theta }-\frac{\Lambda }{e_{A}}%
c_{\varphi }s_{\theta } \\ 
\frac{2e_{B}^{2}\tilde{\Lambda}}{\Lambda ^{2}}c_{\varphi }s_{\theta }-i\frac{%
e_{B}^{2}}{\Lambda ^{2}}c_{\varphi }c_{\theta }-i\frac{\Lambda }{e_{A}}%
s_{\varphi }s_{\theta } \\ 
\frac{4e_{A}e_{B}}{\Lambda ^{2}\sin ^{2}\theta }s_{\theta }^{2}c_{\theta
}^{2}\left( -is_{\varphi }c_{\theta }-2\tilde{\Lambda}s_{\varphi }s_{\theta
}\right)  \\ 
\frac{4e_{A}e_{B}}{\Lambda ^{2}\sin ^{2}\theta }s_{\theta }^{2}c_{\theta
}^{2}\left( c_{\varphi }c_{\theta }+2i\tilde{\Lambda}c_{\varphi }s_{\theta
}\right) 
\end{array}%
\right) 
\end{equation*}%
\begin{equation*}
B_{3}=\left( 
\begin{array}{c}
is_{\varphi }s_{\theta } \\ 
c_{\varphi }s_{\theta } \\ 
-i\frac{2e_{B}\tilde{\Lambda}}{\Lambda }c_{\varphi }s_{\theta }+\frac{e_{B}}{%
\Lambda }c_{\varphi }c_{\theta }+\frac{e_{B}}{e_{A}}s_{\varphi }s_{\theta }
\\ 
\frac{2e_{B}\tilde{\Lambda}}{\Lambda }s_{\varphi }s_{\theta }-i\frac{e_{B}}{%
\Lambda }s_{\varphi }c_{\theta }-i\frac{e_{B}}{e_{A}}c_{\varphi }s_{\theta }%
\end{array}%
\right) \ ,\ C_{3}=\left( 
\begin{array}{c}
\frac{2e_{B}^{2}\tilde{\Lambda}}{\Lambda ^{2}}c_{\varphi }c_{\theta }-i\frac{%
e_{B}^{2}}{\Lambda ^{2}}c_{\varphi }s_{\theta }+i\frac{\Lambda }{e_{A}}%
s_{\varphi }c_{\theta } \\ 
-i\frac{2e_{B}^{2}\tilde{\Lambda}}{\Lambda ^{2}}s_{\varphi }c_{\theta }+%
\frac{e_{B}^{2}}{\Lambda ^{2}}s_{\varphi }s_{\theta }-\frac{\Lambda }{e_{A}}%
c_{\varphi }c_{\theta } \\ 
\frac{4e_{A}e_{B}}{\Lambda ^{2}\sin ^{2}\theta }s_{\theta }^{2}c_{\theta
}^{2}\left( c_{\varphi }s_{\theta }+2i\tilde{\Lambda}c_{\varphi }c_{\theta
}\right)  \\ 
\frac{4e_{A}e_{B}}{\Lambda ^{2}\sin ^{2}\theta }s_{\theta }^{2}c_{\theta
}^{2}\left( is_{\varphi }s_{\theta }+2\tilde{\Lambda}s_{\varphi }c_{\theta
}\right) 
\end{array}%
\right) 
\end{equation*}%
\begin{equation*}
B_{4}=\left( 
\begin{array}{c}
-is_{\varphi }c_{\theta } \\ 
c_{\varphi }c_{\theta } \\ 
-i\frac{2e_{B}\tilde{\Lambda}}{\Lambda }c_{\varphi }c_{\theta }-\frac{e_{B}}{%
\Lambda }c_{\varphi }s_{\theta }-\frac{e_{B}}{e_{A}}s_{\varphi }c_{\theta }
\\ 
-\frac{2e_{B}\tilde{\Lambda}}{\Lambda }s_{\varphi }c_{\theta }-i\frac{e_{B}}{%
\Lambda }s_{\varphi }s_{\theta }-i\frac{e_{B}}{e_{A}}c_{\varphi }c_{\theta }%
\end{array}%
\right) \ ,\ C_{4}=\left( 
\begin{array}{c}
-\frac{2e_{B}^{2}\tilde{\Lambda}}{\Lambda ^{2}}c_{\varphi }s_{\theta }-i%
\frac{e_{B}^{2}}{\Lambda ^{2}}c_{\varphi }c_{\theta }+i\frac{\Lambda }{e_{A}}%
s_{\varphi }s_{\theta } \\ 
-i\frac{2e_{B}^{2}\tilde{\Lambda}}{\Lambda ^{2}}s_{\varphi }s_{\theta }-%
\frac{e_{B}^{2}}{\Lambda ^{2}}s_{\varphi }c_{\theta }+\frac{\Lambda }{e_{A}}%
c_{\varphi }s_{\theta } \\ 
\frac{4e_{A}e_{B}}{\Lambda ^{2}\sin ^{2}\theta }s_{\theta }^{2}c_{\theta
}^{2}\left( c_{\varphi }c_{\theta }-2i\tilde{\Lambda}c_{\varphi }s_{\theta
}\right)  \\ 
\frac{4e_{A}e_{B}}{\Lambda ^{2}\sin ^{2}\theta }s_{\theta }^{2}c_{\theta
}^{2}\left( -is_{\varphi }c_{\theta }+2\tilde{\Lambda}s_{\varphi }s_{\theta
}\right) 
\end{array}%
\right) 
\end{equation*}%
Where $c_{\varphi }=\cos \frac{\varphi }{2},\,s_{\varphi }=\sin \frac{%
\varphi }{2},\,c_{\theta }=\cos \frac{\theta }{2},\,s_{\theta }=\sin \frac{%
\theta }{2}$ .

\end{document}